\documentclass[prd,twocolumn,showpacs,floatfix,amsmath,nofootinbib,amssymb,floatfix]{revtex4}
\usepackage{graphicx,color,dcolumn,booktabs,bm}
\usepackage{longtable,lscape}
\usepackage{txfonts}
\usepackage{overpic}
\usepackage{amssymb}
\usepackage{amsmath}
\usepackage{indentfirst}
\usepackage{feynmf}   
\usepackage{slashed}  
\usepackage{cases}
\usepackage{color}
\usepackage{multirow}
\usepackage{epstopdf}
\usepackage{graphicx,color,dcolumn,booktabs,bm}
\usepackage{ulem}
\usepackage[colorlinks, citecolor=blue,anchorcolor=red,menucolor=red, linkcolor=red,filecolor=red,runcolor=red,urlcolor=blue,frenchlinks=red]{hyperref}

\def\Tr{\textnormal{Tr}}

 \allowdisplaybreaks

\begin{document}

\title{ $D D^{*}$ potentials in chiral effective field theory and possible molecular states }

\author{Hao Xu$^{1,2,3}$}
\author{Bo Wang$^{1,2,4,5}$}
\author{Zhan-Wei Liu$^{1,2}$}\email{liuzhanwei@lzu.edu.cn}
\author{Xiang Liu$^{1,2}$}\email{xiangliu@lzu.edu.cn}
\affiliation{$^1$School of Physical Science and Technology, Lanzhou University,
Lanzhou 730000, China\\
$^2$Research Center for Hadron and CSR Physics,
Lanzhou University $\&$ Institute of Modern Physics of CAS,
Lanzhou 730000, China\\
$^3$Department of Applied Physics, School of Science, Northwestern Polytechnical University, Xi’an 710129, China\\
$^4$ School of Physics and State Key Laboratory of Nuclear Physics and Technology, Peking University, Beijing 100871, China\\
$^5$Center of High Energy Physics, Peking University, Beijing 100871, China
}

\begin{abstract}
The $DD^{*}$ potentials are studied within the framework of heavy meson chiral effective field theory. We have obtained the effective potentials of the $DD^{*}$ system up to $O(\epsilon^2)$ at one loop level. In addition to the one-pion exchange contribution, the contact and two-pion exchange interactions are also investigated in detail. Furthermore, we have searched for the possible molecular states by solving Schr\"odinger equation with the potentials. We notice that the contact and two-pion exchange potentials are non-negligible numerically and important for the existence of a bound state. In our results, no bound state is found in the $I=1$ channel within a wide range of cutoff parameter, while there exists a bound state in the $I=0$ channel as the cutoff is near $m_\rho$ in our approach.

\end{abstract}

\pacs{14.40.Pq, 13.25.Hw}

\maketitle

\section{introduction}\label{sec1}

Chiral effective field theory (ChEFT) is an effective field theory respecting the chiral symmetry of Quantum chromodynamics (QCD) at low momenta. A prominent feature of ChEFT is that the results are expanded as a power series of small momenta rather than small coupling constants, which enables us to systematically study into the non-perturbative regime of the strong interaction. Pseudo-Goldstone bosons such as pion and kaon, with light masses, play very important roles for the low energy processes. Chiral symmetry constrains the form of the interaction quite strongly. Owing to the clear power counting scheme, ChEFT is very powerful to investigate the properties of light pseudoscalar bosons \cite{Weinberg:1978kz,Gasser:1983yg,Scherer:2012tk}.

The situation becomes complicated when heavy hadrons involve. The power counting rule is broken because of large hadron masses. However, for the system with single heavy hadron and few light pseudoscalar bosons, the power counting scheme can be easily rebuilt, and many approaches of ChEFT have been developed to deal with the relevant scattering, interaction, electromagnetic moments, and other properties of such system. Heavy hadron chiral perturbation theory, the infrared regularization, and the extended-on-mass-shell scheme are frequently used in one heavy hadron sector \cite{Jenkins:1990jv,Bernard:1992qa,Bernard:1995dp,Becher:1999he,Fuchs:2003qc,Geng:2013xn,Guo:2009ct,Liu:2009uz,Geng:2010vw,Wang:2012bu,Altenbuchinger:2013vwa,Guo:2015dha,Yao:2015qia,Du:2017ttu,Du:2016xbh}. Unfortunately, these approaches cannot be directly extended to study the properties about few heavy hadrons, like nuclear force.

Two-nucleon interaction bears another power counting problem. Two approximately on-shell nucleons in loop diagrams cause extra enhancement compared to the naive power counting, which prevents us from calculating scattering matrix directly.
Weinberg proposed a framework to deal with the issue \cite{Weinberg:1990rz,Weinberg:1991um}. One can first calculate an effective potential, i.e., sum of all two-particle irreducible (2PI) diagrams, and then iterate it with equations, such as Lippmann-Schwinger and Schr\"odinger equation, to retrieve two-particle reducible (2PR) contributions. The Weinberg's formalism has been further extended and developed \cite{Ordonez:1992xp,Ordonez:1995rz,Epelbaum:1998ka,Epelbaum:1999dj,Long:2011qx,Long:2016vnq,Epelbaum:2014efa,Kang:2013uia,Dai:2017ont,Ren:2016jna,Epelbaum:2008ga,Machleidt:2011zz}. For example, a unitary transformation is presented to remove the energy dependence of the potential in Refs. \cite{Epelbaum:1998ka,Epelbaum:1999dj}. The renormalization of potentials are carefully studied in Refs.~\cite{Valderrama:2009ei,Valderrama:2011mv,Long:2011qx,Long:2012ve,Long:2016vnq}. The authors in Ref.~\cite{Epelbaum:2014efa} revisit the nucleon-nucleon
potential up to NNNLO within ChEFT. In Refs.~\cite{Kang:2013uia,Dai:2017ont}, the nucleon-antinucleon
potential is investigated within ChEFT. Very recently, a covariant formalism of the $N$-$N$ interaction is proposed in Ref.~\cite{Ren:2016jna}. Three body and even four body nuclear forces have been systematically studied within ChEFT, see Refs. \cite{Epelbaum:2008ga,Machleidt:2011zz} for a review.  The application of ChEFT has been definitely advancing our understanding of the nuclear force \cite{Meissner:2015wva}.

With successes in the study of nuclear force, one may wonder whether ChEFT can help us to comprehend the interactions of heavy (charmed, bottomed) meson systems. Obviously, since heavy meson is heavier, we can make some assumptions such as the heavy quark limit without worries, and thus heavy hadron ChEFT is even more suitable than that in the nucleon system.

The $XYZ$ and similar exotic states have attracted a lot of interest in the hadron physics, and it is well known the interaction between heavy mesons is quite responsible for the strange behavior at close threshold in charmonium and bottomonium spectra (see Ref.~\cite{Chen:2016qju} for a review). This starts with the discovery of the famous $X(3872)$, which was observed by the Belle Collaboration in $B$ decay process $B^{+-}\to K^{+-} \pi^+ \pi^- J / \psi$ in 2003 \cite{Choi:2003ue}. $X(3872)$ is extremely close to the threshold of $D^0\bar{D}^{*0}$. Its mass is much smaller than quark model (such as the Godfrey-Isgur model \cite{Godfrey:1985xj}) predictions if it is regarded as $\chi^\prime_{c1}(2P) $ charmonium, and moreover it has a large decay width for the isospin violation process $X(3872)\to J/\psi \rho$. After that, more and more $XYZ$ and other exotic states candidates were discovered, such as recent observed pentaquark $P_c(4380)^+$ and $P_c(4450)^+$ \cite{Aaij:2015tga} and still debated $X(5568)$ \cite{D0:2016mwd}.

There are many models dealing with these states, such as the one-boson-exchange molecular model, some underlying multiquark models, kinematical effect, and so on (see the review \cite{Chen:2016qju}). For example, in Refs.~\cite{Molina:2009ct,Aceti:2014uea} $D^*_{(s)}\bar{D}^*_{(s)}$ and $D\bar{D}^*$ systems are studied within
the local hidden gauge formalism to dynamically generate $Y(3940)$, $Z(3930)$, $X(4160)$ and $Z_c(3900)$. In
Ref.~\cite{He:2015mja}, the authors have investigated the $D\bar{D}^*$ system and its relation to $Z_c(3900)$ using the covariant
spectator theory. $Z_c(3900)$ is also studied from the pole counting rule \cite{Gong:2016hlt}. The authors in Ref.~\cite{Yang:2017prf} have discussed $D^{(*)}\bar{D}^{(*)}$ with the constituent quark models, and solved the four-body Schr\"{o}dinger equation with the Gaussian expansion method. The contact interaction of $D\bar{D}^*$($B\bar{B}^*$) is specially investigated in Ref.~\cite{AlFiky:2005jd} with the effective field theory, which is implemented with the heavy quark symmetry. The $D\bar{D}^\ast$ system is also intensively studied with different kinds of effective field theories, see Refs.~\cite{Fleming:2007rp,Fleming:2008yn,Baru:2011rs,Valderrama:2012jv,Nieves:2012tt,Baru:2015nea,Meng:2014ota,Baru:2015tfa,Braaten:2015tga,Jansen:2015lha,Baru:2016iwj} and many other works citied therein. For example, in Ref.~\cite{Fleming:2007rp}, the authors studied the $D\bar{D}^{*}$ with XEFT using perturbative pions. In Ref~\cite{Baru:2011rs}, the authors studied $X(3872)$ and $D\bar{D}^{*}$ using non-perturbative pions. Moreover, the authors in Ref.~\cite{Baru:2016iwj} further included the effects of the $D^*$ width. In Ref.~\cite{Guo:2017jvc}, the study of hadronic molecules with effective field theories are reviewed.

As mentioned above, there are many models dealing with heavy meson systems. Among them, the one-boson-exchange model has interpreted many exotic phenomenas and made some predictions which have been verified by the later discoveries of new particles at experiment. This model can provide the dynamical potentials of hadron systems, and then one can solve the Schr\"odinger equation to see if there is a bound state.
The model has been widely used to study the interaction of the two-heavy-hadron systems and related exotic states. The research of the charmed-anticharmed system and $X(3872)$ experiences a long progress. It starts from the pion and $\sigma$ exchanges in early Ref.~\cite{Liu:2008fh}, directly extends to the multi-state exchanges  \cite{Liu:2008tn}, and then includes more complicated effects from $S$-$D$ mixing \cite{Lee:2009hy}, isospin violation \cite{Li:2012cs}, and so on. For the investigation of nuclear force, after the boson exchange model develops for decades (see the discussion in Ref.~\cite{Epelbaum:2001fm}), ChEFT is applied at last and help us systematically build the modern system of knowledge. Following their steps, it is natural to introduce ChEFT into the study of heavy meson systems after the one-boson-exchange model.

There exist many works on heavy mesons system with one-boson-exchange model and effective field theories, as mentioned above. It is interesting to investigate their higher order effects in chiral effective field theory, and then discuss the potential in coordinate space and search for the bound state by solving Schr\"odinger equation. We will also compare the results with one-boson-exchange model.

In this work, we focus on the doubly charmed-meson system $DD^{*}$, which is clearer than the hidden charmed system for the absence of annihilation channels. It provides us another insight to understand the heavy-flavor dynamics and non-perturbative QCD. Furthermore, it is analogous to deuteron since they both have contact, one-pion exchange (OPE), and two-pion exchange (TPE) contributions without annihilation channels in our framework.

Till now, the only observed doubly heavy-flavor system is the $\Xi_{cc}^{++}$ baryon which was first discovered by SELEX collaboration \cite{Mattson:2002vu}. Systems like $ccu$ and $ccd$ have been discussed a lot, and their properties such as masses and electromagnetic moments still need more efforts to get clarified \cite{Sun:2014aya,Yu:2017zst,Hu:2005gf,Chen:2016spr,Chen:2017sbg,Li:2017cfz,Wang:2017azm,Karliner:2017elp}.
Very recently, LHCb group confirmed the existence of $\Xi_{cc}^{++}$ but disfavored the mass measured at SELEX \cite{Aaij:2017ueg}.   With the technique and apparatus well developed nowadays, it is also possible to search for the doubly charmed boson made of $DD^{*}$ at experiment.

In Ref.~\cite{Ohkoda:2012hv}, the authors studied $D^{(*)}D^{(*)}$($B^{(*)}B^{(*)}$) system to search for bound and resonant states, and they used pion and vector meson exchange potentials which are constrained by heavy quark symmetry and chiral symmetry. They found that in isospin 0 channel there exists a bound state in $S$-wave with binding energy $62.3$ MeV, and no bound state is found in isospin 1 with $S$-wave. In Ref.~\cite{Li:2012ss} the authors have studied $D^{(*)}D^{(*)}$ system using the one-boson-exchange model, and found that
there exists a bound state consisting of $DD^{*}$ with binding energy $5\sim43$ MeV in the isospin 0 channel. The authors in Ref.~\cite{Abreu:2015jma} investigate deuteron-like molecules with both open charm and bottom using the heavy-meson effective theory. In Ref.~\cite{Sakai:2017avl},
charm-beauty meson bound states are dynamically generated from the $B^{(*)}D^{(*)}$ and $B^{(*)}\bar{D}^{(*)}$ interactions, and
they also give the informations of the scattering lengths. There also exist lattice studies on $BB$ and $BB^*$ interaction \cite{Detmold:2007wk,Bicudo:2015kna,Francis:2016hui}. Especially in Ref.~\cite{Francis:2016hui}, the authors considered both diquark-antidiquark and meson-meson configuration. In Ref.~\cite{Liu:2012vd}, we have investigated $\bar{B}\bar{B}$ interaction within heavy meson chiral effective field theory (HMChEFT). We obtain the potentials of the $\bar{B}\bar{B}$ system at one loop level, and have discussed the contact and two-pion exchange contributions in momentum space.

We investigate the $DD^{*}$ system in this work. As we mentioned before, we need to study the potentials first, and then access physical observables indirectly. Furthermore, the potential in coordinate space can give us more intuitive information about interaction between mesons, and we can further solve a dynamic equation to see whether there exists a hadronic molecule. This paper is organized as follows. After introduction, we elucidate the framework in Sec.~\ref{sec2}. In Sec.~\ref{sec3}, we give results of potentials in momentum space. In Sec.~\ref{sec4}, we study the potential in coordinate space to search possible molecules. At last, we summarize our conclusions.

\section{Lagrangians and Weinberg scheme} \label{sec2}
To study the $D D^{*} $ system under HMChEFT, we need to show Lagrangians and provide results systematically in a strict power counting scheme. Our results are arranged order by order with the small parameter $\epsilon=p/\Lambda_\chi$, where $p$ can be the momentum of pion, the residual momentum of heavy mesons, or the $D$-$D^*$ mass splitting, and $\Lambda_\chi$ represents either the chiral breaking scale or the mass of the heavy mesons. In this work, flavor SU(2) symmetry is always kept.

\subsection{Lagragians at the leading order}
At the leading order $O(\epsilon^0)$, both OPE diagrams and contact diagrams contribute to the amplitudes, and thus we should first build the Langrangians for $DD^*\pi$ interaction vertices, the corresponding contact vertices, and so on.

The $DD^*\pi$ Lagrangian at leading order \cite{Burdman:1992gh,Wise:1992hn,Yan:1992gz} is given by 
  \begin{eqnarray}\label{LagrangianHpi1}
  \mathcal L^{(1)}_{H\phi}&=&-\langle (i v\cdot \partial H)\bar H
  \rangle
  +\langle H v\cdot \Gamma \bar H \rangle
  +g\langle H \slashed u \gamma_5 \bar H\rangle
  \nonumber \\  &&-\frac18 \delta \langle H \sigma^{\mu\nu} \bar H \sigma_{\mu\nu} \rangle.
  \end{eqnarray}
 In the above, $H$ field represents the $(D,D^*)$ doublet in the heavy quark limit
  \begin{eqnarray} \label{Hfield}
  && H=\frac{1+\slashed v}{2}\left(P^*_\mu\gamma^\mu+iP\gamma_5\right),\quad \nonumber \\
  &&\bar H=\gamma^0 H^\dag \gamma^0 = \left(P^{*\dag}_\mu \gamma^\mu+iP^\dag \gamma_5\right) \frac{1+\slashed v}{2},\nonumber\\
  && P=(D^0, D^+), \quad P^*_\mu=(D^{*0}, D^{*+})_\mu.
  \end{eqnarray}
 $v=(1,0,0,0)$ stands for the 4-velocity of the $H$ field. The last term in Eq.(1) is included to account for $D$-$D^*$ mass shift which is not zero in the chiral limit, and $\delta$ is the mass difference in
 $(D,D^*)$ doublet. The axial vector field $u$ and chiral connection $\Gamma$ are expressed as
  \begin{equation}
  \Gamma_\mu = {i\over 2} [\xi^\dagger, \partial_\mu\xi],\quad
  u_\mu={i\over 2} \{\xi^\dagger, \partial_\mu \xi\},
  \end{equation}
 where $\xi =\exp(i \phi/2f)$, $f$ is the bare constant for pion decay, and
  \begin{equation}
  \phi=\sqrt2\left(
  \begin{array}{cc}
  \frac{\pi^0}{\sqrt2}&\pi^+\\
  \pi^-&-\frac{\pi^0}{\sqrt2}\\
  \end{array}\right).
  \end{equation}

 The contact Lagrangian at $O(\epsilon^0)$ is constructed as follows \cite{AlFiky:2005jd,Valderrama:2012jv,Liu:2012vd}
  \begin{eqnarray}\label{Lagrangian4H0}
  \mathcal L^{(0)}_{4H}&=&D_{a} \textnormal{Tr} [H \gamma_\mu \bar H ] \textnormal{Tr} [ H
  \gamma^\mu\bar H] \nonumber \\  &&+D_{b} \textnormal{Tr} [H \gamma_\mu\gamma_5 \bar H ] \textnormal{Tr} [ H
  \gamma^\mu\gamma_5\bar H] \nonumber\\  && +E_{a} \textnormal{Tr} [H
  \gamma_\mu\tau^a \bar H ] \textnormal{Tr} [ H \gamma^\mu\tau_a\bar H] \nonumber\\
  &&+E_{b} \textnormal{Tr} [H \gamma_\mu\gamma_5\tau^a \bar H ] \textnormal{Tr} [ H
  \gamma^\mu\gamma_5\tau_a\bar H],
  \end{eqnarray}
 where $D_a$, $D_b$, $E_a$, $E_b$ are four independent low energy constants (LECs).

\subsection{Lagrangians at the next to leading order}
 At chiral order $O(\epsilon^2)$, the total amplitudes consists of the
 contact corrections, OPE corrections, and TPE amplitudes.
 These one-loop amplitudes must be renormalized with the help of $O(\epsilon^2)$ Lagrangians. The divergences
  in the one-loop amplitudes are canceled
 by the infinite parts of the LECs in the following lagrangians \cite{Liu:2012vd},
 \begin{eqnarray}\label{Lagrangian4H21}
 \mathcal L^{(2,h)}_{4H}&=& D_{a}^h\Tr[H \gamma_\mu \bar H ]\Tr[ H
 \gamma^\mu\bar H] \Tr(\chi_+) \nonumber \\ &&+D_{b}^h\Tr[H \gamma_\mu\gamma_5
 \bar H ]\Tr[ H \gamma^\mu\gamma_5\bar H]\Tr(\chi_+) \nonumber\\&&
 +E_{a}^h \Tr[H \gamma_\mu\tau^a \bar H ]\Tr[ H
 \gamma^\mu\tau_a\bar H]\Tr(\chi_+)  \nonumber \\ &&+E_{b}^h \Tr[H
 \gamma_\mu\gamma_5\tau^a \bar H ]\Tr[ H
 \gamma^\mu\gamma_5\tau_a\bar H]\Tr(\chi_+),  \label{Lagrangian4H2h}
 \\
 \mathcal L^{(2,v)}_{4H}&=& \{D_{a1}^v\Tr[(v\cdot D H) \gamma_\mu
 (v\cdot D \bar H) ]\Tr[ H \gamma^\mu\bar H] \nonumber \\ &&+D_{a2}^v\Tr[(v\cdot D
 H) \gamma_\mu \bar H ]\Tr[ (v\cdot D H) \gamma^\mu\bar H]
 \nonumber\\&& +D_{a3}^v\Tr[(v\cdot D H) \gamma_\mu \bar H ]\Tr[  H
 \gamma^\mu(v\cdot D \bar H)] +\nonumber \\ && D_{a4}^v\Tr[((v\cdot D)^2 H)
 \gamma_\mu \bar H ]\Tr[  H \gamma^\mu\bar H ] \nonumber\\&&
 +D_{b1}^v\Tr[(v\cdot D H) \gamma_\mu\gamma_5 (v\cdot D \bar H)
 ]\Tr[ H \gamma^\mu\gamma_5\bar H]+... \nonumber \\ && +E_{a1}^v \Tr[(v\cdot D H)
 \gamma_\mu\tau^a (v\cdot D \bar H) ]\Tr[ H
 \gamma^\mu\tau_a\bar H]+... \nonumber\\&& +E_{b1}^v\Tr[(v\cdot
 D H) \gamma_\mu\gamma_5\tau^a (v\cdot D \bar H) ]\Tr[ H
 \gamma^\mu\gamma_5\tau_a \bar H] \nonumber \\ && +...\} +\text{H.c.},
\label{Lagrangian4H2v}
 \\
 \mathcal L^{(2,q)}_{4H}&=& \{D_1^q\Tr[(D^\mu H) \gamma_\mu\gamma_5
 (D^\nu \bar H) ]\Tr[ H \gamma_\nu\gamma_5\bar H] \nonumber \\ && +D_2^q\Tr[(D^\mu
 H) \gamma_\mu\gamma_5 \bar H ]\Tr[ (D^\nu H)
 \gamma_\nu\gamma_5\bar H] \nonumber\\&& +D_3^q\Tr[(D^\mu H)
 \gamma_\mu\gamma_5 \bar H ]\Tr[ H \gamma_\nu\gamma_5(D^\nu \bar
 H)] \nonumber \\ && +D_4^q\Tr[(D^\mu D^\nu H) \gamma_\mu\gamma_5 \bar H ]\Tr[ H
 \gamma_\nu\gamma_5 \bar H] \nonumber\\&& +E_1^q\Tr[(D^\mu H)
 \gamma_\mu\gamma_5 \tau^a(D^\nu \bar H) ]\Tr[ H
 \gamma_\nu\gamma_5\tau_a\bar H] \nonumber \\ && +...\}+\text{H.c.},
  \cdots,  \label{Lagrangian4H24}
 \end{eqnarray}
 where
 \begin{eqnarray}
 \tilde\chi_\pm&=&\chi_\pm-\frac12\Tr[\chi_\pm], \nonumber\\
 \chi_\pm&=&\xi^\dagger\chi\xi^\dagger\pm\xi\chi\xi , \nonumber\\
 \chi&=&m_\pi^2.
 \end{eqnarray}
Note that, the term $\mathcal L^{(2,d)}_{4H}$ in Ref.~\cite{Liu:2012vd} vanishes in our SU(2) case.

In addition to canceling the divergences of the loop diagrams, the above Lagrangians also contain finite
 parts that contribute to tree-level diagrams at $O(\epsilon^2)$. They are governed by a large amount
 of LECs appearing in Eqs.~(\ref{Lagrangian4H21})-(\ref{Lagrangian4H24}).

\subsection{Weinberg scheme}
\begin{figure}[htpb]
 	\begin{center}
 		\includegraphics[scale=0.66]{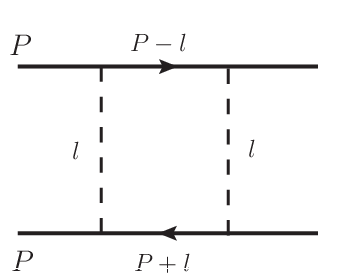}
 		\caption{A typical TPE box diagram of the nucleon-nucleon
 			interaction. The solid line stands for the nucleon and the dashed line stands for
 		the	pion.}\label{2piExchangeBoxDiagram}
 	\end{center}
  \end{figure}

 In this work, we adopt the power counting scheme from Weinberg to study the $D D^{*}$ systems \cite{Weinberg:1990rz,Weinberg:1991um}.
This framework has been widely applied to nucleon-nucleon system as mentioned in the introduction.
 Let us start with a nucleon-nucleon TPE box diagram depicted in Fig.~\ref{2piExchangeBoxDiagram}.
 As illustrated in Ref.~\cite{Liu:2012vd}, the amplitude can be written under the heavy hadron
 formalism:
   \begin{eqnarray} \label{BoxDiagramIntegral}
   && i\int d^4l \frac{i}{l^0+P^0-\frac{\vec{q}_1^2}{2M_N} +i \varepsilon}  \frac{i}{-l^0+P^0-\frac{\vec{q}^2_2}{2M_N}+i \varepsilon} \times \cdots \nonumber \\
   &&= i\int d^3l \int dl^0 \frac{i}{l^0+P^0-\frac{\vec{q}^2_1}{2M_N}+i \varepsilon} \frac{i}{-l^0+P^0-\frac{\vec{q}^2_2}{2M_N}+i \varepsilon} \nonumber \\
   && \quad \times \cdots \nonumber \\
   &&=\int d^3l \frac{\pi}{P^0-\frac{1}{2} \left( \frac{ \vec{q}^2_1}{2M_N}+\frac{\vec{q}^2_2}{2M_N} \right)+i \varepsilon} \cdots \nonumber \\
   &&=\int d^3l \frac{\pi}{\frac{\vec{P}^2}{(2M_N)}-\frac{1}{2} \left( \frac{ \vec{q}^2_1}{2M_N}+\frac{\vec{q}^2_2}{2M_N} \right)+i \varepsilon} \cdots \nonumber \\
   &&=-\int d^3l \frac{\pi}{\frac{\vec{l}^2}{(2M_N)}+i \varepsilon} \cdots,
   \end{eqnarray}
  where $m_N$ is the mass of the nucleon, $\vec{q_1}=\vec{P}+\vec{l}$ and $\vec{q_2}=\vec{P}-\vec{l}$.
   Naive power counting gives the $l^0$ integral $O(|\vec{P}|^{-1})$, while we notice from
  Eq.~(\ref{BoxDiagramIntegral}) that the $l^0$ integral should be of $O(|\vec{P}|^{-2})$, i.e. the
  true order is enhanced by $|\vec{P}|^{-1}$. Such enhancement definitely violates the power counting
  rule, which would invalidate the chiral expansion. As pointed out in Ref.~\cite{Weinberg:1990rz,Weinberg:1991um}
  , the origin of such a contradiction comes from
   double poles in Eq.~(\ref{BoxDiagramIntegral}) which relates to two-particle-reducible (2PR) part
  of the box diagram in Fig.~\ref{2piExchangeBoxDiagram}.

 With the above analysis in mind, we just fall into the same situation when studying the interaction of the
 doubly-charmed meson pair, and thus can not directly calculate the scattering amplitude.
 Alternatively, we apply the Weinberg's power counting scheme. First, with the usual power counting
 rule, we compute the 2PI contributions of all diagrams, and this leads
 to effective potentials. Then we substitute the potentials into iterated equations such as
 Lippmann-Schwinger equation or Schr\"{o}dinger equation to recover the 2PR contributions. Finally,
 we would obtain the desired scattering amplitudes or energy levels.

\section{Effective potentials of $DD^{*}$ system}\label{sec3}

The effective potentials of $DD^{*}$ system receive contributions from the contact and OPE diagrams at the leading order $O(\epsilon^0)$. At the next to leading order $O(\epsilon^2)$, there are both tree and one-loop corrections. The effective potentials $\mathcal{V}$ are related to the Feynman amplitudes $\mathcal{M}$ of 2PI diagrams
\begin{eqnarray}
\mathcal{V} = \frac{-1}{4} \mathcal{M},
\end{eqnarray}
which follows from the one-boson-exchange model despite some differences
in conventions \cite{Sun:2012zzd,Sun:2011uh}.

\begin{figure}[htpb]
  	\begin{center}
  		\includegraphics[scale=0.45]{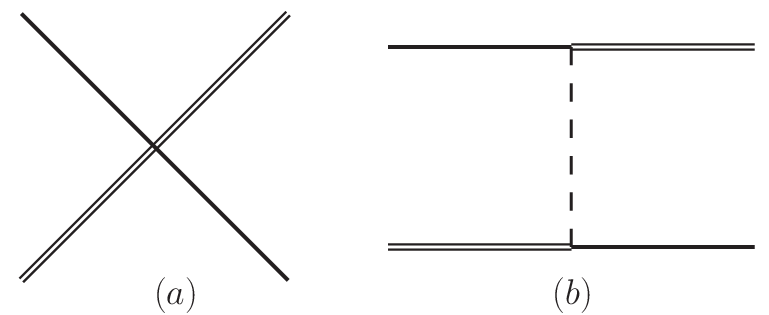}
  		\caption{Tree-level digrams of the processes $DD^{*}\to DD^{*}$
  	    at $O(\epsilon^0)$. The left diagram relates to the contact terms, and the right one is the one-pion-exchange diagram. The solid, double-solid, and dashed lines stand for $D$, $D^{*}$, pion, respectively.}\label{O0TreeDiagram}
  	\end{center}
  \end{figure}

At the lowest order $O(\epsilon^0)$, there are two
diagrams at tree level illustrated in Fig.~\ref{O0TreeDiagram}. They represent the contact and OPE contributions, individually. The contact terms mainly affect the short range interaction between particles while the OPE contribution determines the behavior of the long range interaction.

With Lagrangians (\ref{LagrangianHpi1}) and (\ref{Lagrangian4H0}), the corresponding amplitudes can be easily computed. For the process $D(p_1)D^{*}(p_2)\to D(p_3)D^{*}(p_4)$ with isospin $I=1$, the amplitudes for diagrams (a) and (b) in Fig.~\ref{O0TreeDiagram} read
\begin{eqnarray}\label{O0ContactAmplitude}
\mathcal{M}^{(0)}_{I=1(a)} &=& i (-8D_a+8D_b-8E_a+8E_b)\varepsilon(p_2)\cdot\varepsilon^*(p_4), \\
\mathcal{M}^{(0)}_{I=1(b)} &=& i(-1) \frac{g^2}{f^2} \frac{p_\mu p_\nu}{p^2-m^2} \varepsilon^\mu(p_2)\varepsilon^{*\nu}(p_4).\label{O11piAmplitude}
\end{eqnarray}
For the process $D(p_1)D^{*}(p_2)\to D(p_3) D^{*}(p_4)$ with $I=0$, the amplitudes are
\begin{eqnarray}
\mathcal{M}^{(0)}_{I=0(a)} &=& i (24E_a+24E_b-8D_a-8D_b)\varepsilon(p_2)\cdot\varepsilon^*(p_4), \\
\mathcal{M}^{(0)}_{I=0(b)} &=& i (-3)\frac{g^2}{f^2} \frac{p_\mu p_\nu}{p^2-m^2} \varepsilon^\mu(p_2)\varepsilon^{*\nu}(p_4).
\label{O01piAmplitude}
\end{eqnarray}
In above equations, momentum $p=p_1-p_4$, the superscript $(0)$ denotes the order $O(\epsilon^0)$, and the subscripts ``$I=0, 1$'' stand for
the process $DD^{*}\to DD^{*}$ with isospin $0,1$, respectively.

At $O(\epsilon^2)$, a number of diagrams emerge. The tree diagrams at $O(\epsilon^2)$ are similar to Fig.~\ref{O0TreeDiagram}~(a), but the vertices should be replaced with those from Lagrangians (\ref{Lagrangian4H2h}-\ref{Lagrangian4H24}). There are additionally three sets of one-loop diagrams.

\begin{figure}[htpb]
  	\begin{center}
  		\includegraphics[scale=0.38]{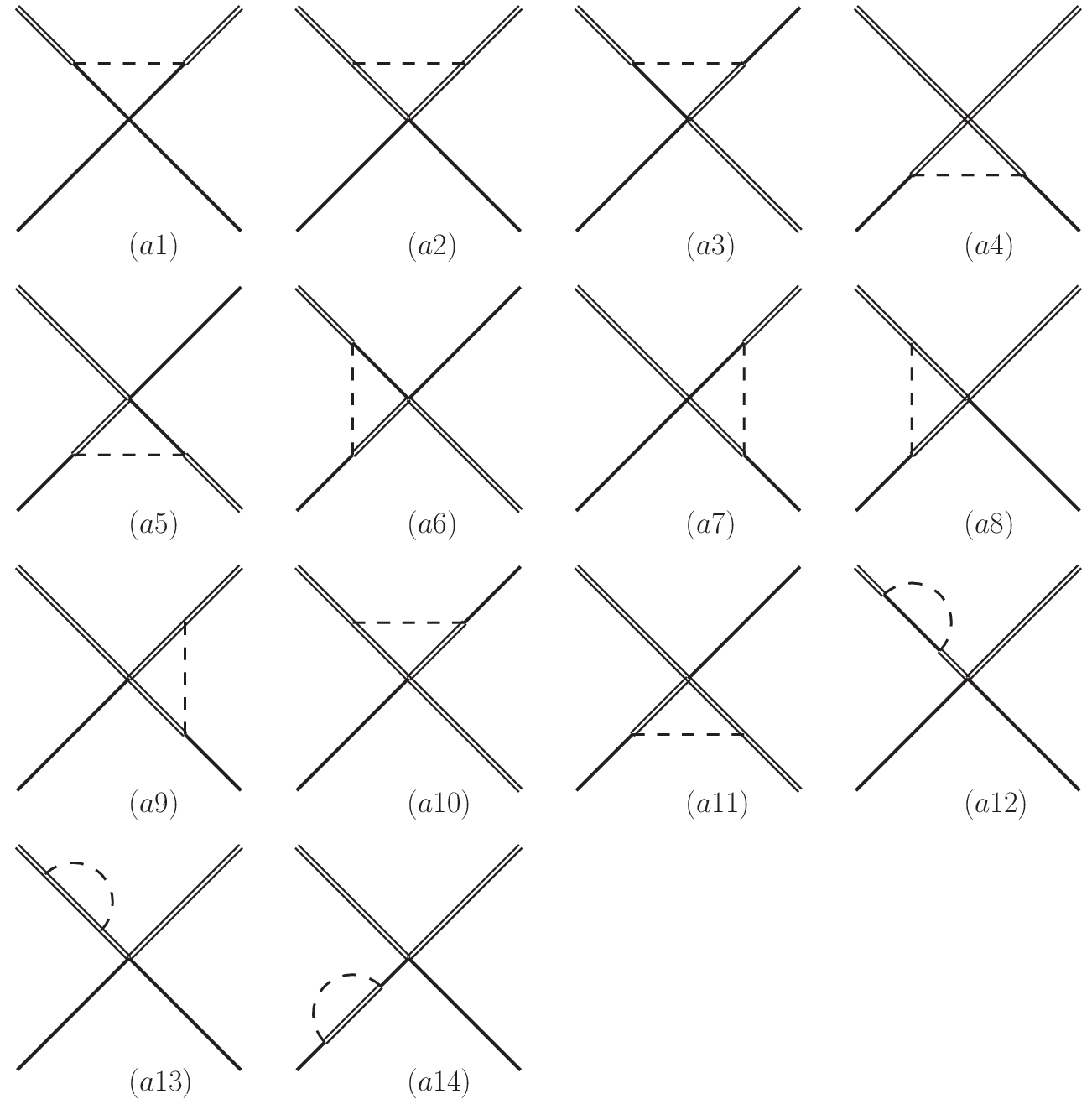}
  		\caption{One-loop corrections to the contact
  			 terms at $O(\epsilon^2)$. The solid, double-solid, and dashed lines stand for $D$, $D^{*}$, pion, respectively.}\label{O2ContactDiagram}
  	\end{center}
  \end{figure}

The diagrams in the first set are for one-loop corrections to the contact terms. They are depicted in Fig.~\ref{O2ContactDiagram}. Diagrams (a12)-(a12) represents contributions from the wave function renormalization of external legs.

\begin{figure}[htpb]
   	\begin{center}
   		\includegraphics[scale=0.43]{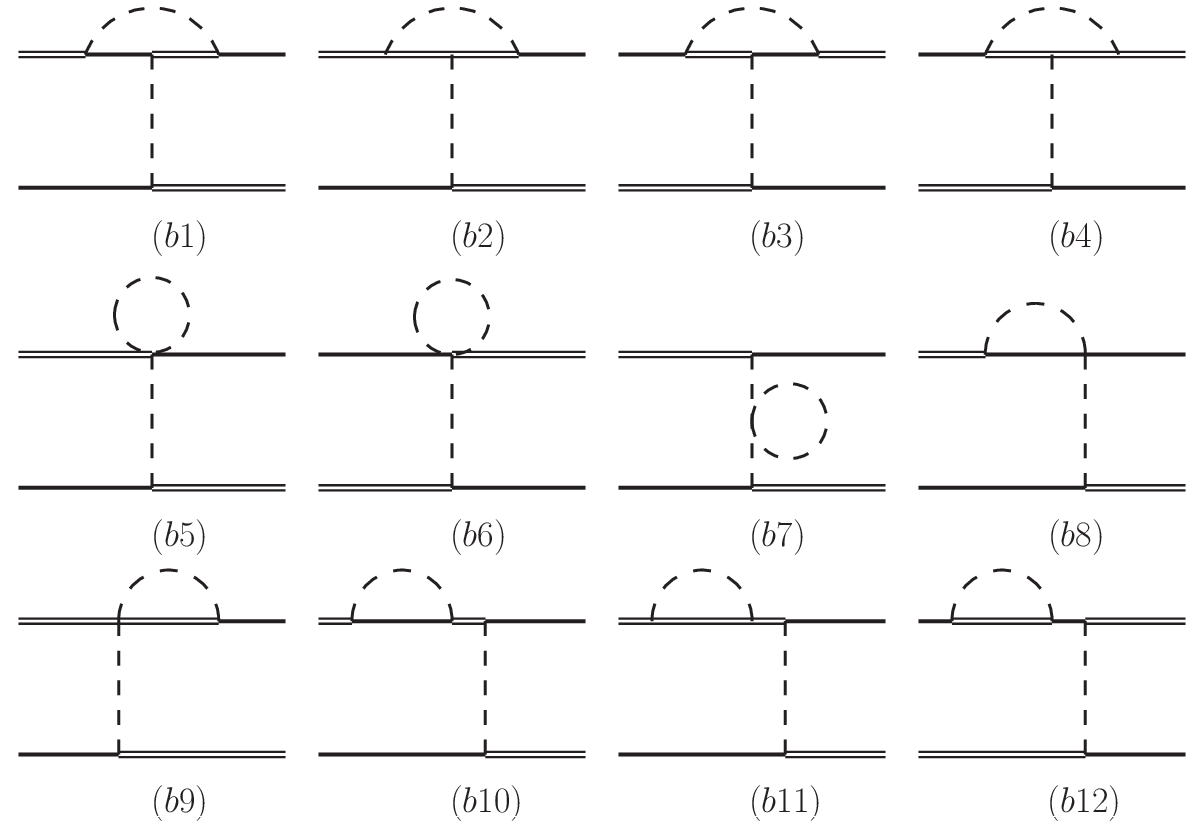}
   		\caption{ One-loop corrections to the one-pion-exchange
   			 diagrams at $O(\epsilon^2)$. The solid, double-solid, and dashed lines stand for $D$, $D^{*}$, pion, respectively.}\label{O21piDiagram}
   	\end{center}
   \end{figure}

We show the second set of diagrams in Fig.~\ref{O21piDiagram}. They represent one-loop corrections to the OPE diagrams. The diagrams~(b1)-(b6) and (b8)-(b9) in Fig.~\ref{O21piDiagram} contribute to the renormalization of the $DD^*\pi$ vertex. Therefore, we must use the value for the bare coupling $g$ in $\mathcal M^{(0)}$ at $O(\epsilon^0)$ to avoid double counting. We show the relation between the bare coupling $g$ and the experiment coupling $g^{(2)}$ in Eq. (\ref{gCorrection}) in Appendix \ref{secAppRBC}. Similarly, the bare decay constant $f$ should be used in Eqs.~(\ref{O11piAmplitude},\ref{O01piAmplitude}), too.

\begin{figure}[htpb]
   	\begin{center}
   		\includegraphics[scale=0.35]{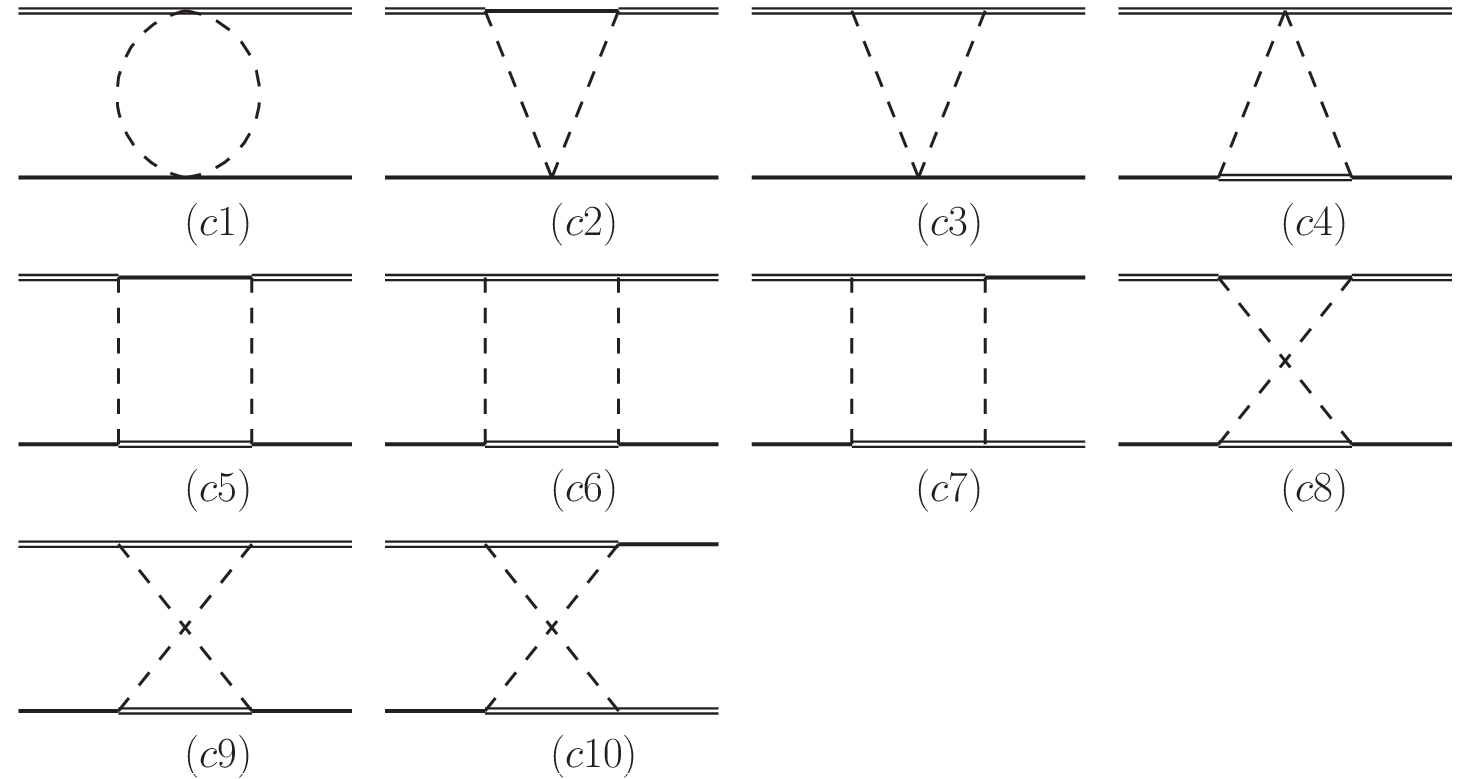}
   		\caption{Two-pion-exchange diagrams at $O(\epsilon^2)$. The solid, double-solid, and dashed lines stand for $D$, $D^{*}$, pion, respectively.}\label{O22piDiagram}
   	\end{center}
   \end{figure}

Final set is for the TPE diagrams which are illustrated in Fig.~\ref{O22piDiagram}. They are important for the medium range interaction.

As discussed in the previous section,
some diagrams, such as the box diagrams in Fig.~\ref{O22piDiagram}, contain a 2PR part that should be subtracted. If there exists a loop function of a box diagram like
   \begin{eqnarray} \label{DiagramWith2PR}
   && \int d^4l \frac{1}{v \cdot l+a+i \varepsilon} \frac{1}{-v \cdot l-a+i \varepsilon} \times \cdots,
   \end{eqnarray}
 following Ref.~\cite{Zhu:2004vw,Liu:2012vd}, we can separate the 2PR and 2PI parts by
   \begin{eqnarray} \label{SubtractionOf2PR}
   && \frac{1}{v \cdot l+a+i \varepsilon} \frac{1}{-v \cdot l-a+i \varepsilon} \nonumber \\
   &&= \frac{1}{v \cdot l+a+i \varepsilon} \left[ -\frac{1}{v \cdot l+a+i \varepsilon}
   +2\pi \delta(v \cdot l + a) \right] .
   \end{eqnarray}
The term proportional to Dirac $\delta$ function is just the 2PR part which should be dropped in potentials.

All the one-loop amplitudes of the diagrams Figs.~\ref{O2ContactDiagram}-\ref{O22piDiagram}
for the processes $DD^{*}\to DD^{*}$
 are shown in Appendix \ref{SecAppA}. The divergences of the loop functions are regularized with the
 dimensional regularization, and subtracted by the modified minimal subtraction scheme.
 Also, we list the definitions of the loop functions in Appendix \ref{SecAppB}.
The finite parts of the high order lagrangians should also contribute to tree level diagrams at $O(\epsilon^2)$, and they
are governed by a large number of LECs. However, it needs plenty of data for $D\bar{D}^*$ (or other channels such as $D\bar{D}$, $DD$) scattering in different partial waves to fit these LECs, but there is still lack now. Therefore in the present work, we only focus on the loop contributions at $O(\epsilon^2)$.

We can easily obtain the potentials $\mathcal{V}_{I=1}^{DD^*}$ and $\mathcal{V}_{I=0}^{DD^*}$ from the Feynman amplitudes by multiplying a factor $-1/4$. 
The polarized vectors in the potentials are delicately dealt with in Ref. \cite{Gulmez:2016scm}.
In this work, we only consider the $S$-wave interaction, which leads to the following substitutions
in Eqs.~(\ref{O0ContactAmplitude})-(\ref{O01piAmplitude}), (\ref{AllAmplitude1})-(\ref{AllAmplitude2})
\begin{eqnarray}
&&\vec{\varepsilon}(p_2)\cdot\vec{\varepsilon}^*(p_4) \rightarrowtail 1, \\
&& \vec{\varepsilon}(p_2)\cdot \vec{p} \; \vec{\varepsilon}^*(p_4) \cdot \vec{p} \rightarrowtail \frac{1}{3} \vec{p}^2,
\end{eqnarray}
where we follow from the one-boson exchange model in Ref. \cite{Sun:2012zzd,Sun:2011uh}.
 After all these
 procedures, the effective potentials $\mathcal{V}_{I=1}^{DD^*}$ and $\mathcal{V}_{I=0}^{DD^*}$ in
the momentum space can be obtained. But, the potentials are energy dependent. A solution to this
problem is proposed in Refs.~\cite{Epelbaum:1998ka,Epelbaum:1999dj}, where they apply
a unitary transformation to get rid of the energy dependence. While in this work, we just take the
transfered energies equal to zero, i.e. $p^0=0$ and $q^0=0$ for simplicity, as in the one-boson exchange
model \cite{Liu:2008fh}.
Also, we take the residual energies of the heavy mesons equal to zero, too.

\section{Numerical Results of potentials in momentum space}\label{sec4}
We display input parameters for the numerical results: $m_\pi=0.139$ GeV, the mass difference $\delta=0.142$ GeV, $f_\pi=0.086$ GeV, and renormalization scale $\mu=4\pi f$. There are many works investigating the constant for $D^{(*)}D^{(*)}\pi$ coupling such as the lattice
study \cite{Abada:2002xe,Becirevic:2012pf,Can:2012tx}, QCD sum rule \cite{Colangelo:1994es,Belyaev:1994zk,Dosch:1995kw,Colangelo:1997rp,Bracco:2011pg}, and other approaches \cite{Colangelo:1994jc,Becirevic:1999fr,ElBennich:2010ha}. The experimental process $D^*\to D\pi$ is fit to obtain the renormalized coupling $g^{(2)}$ \cite{Olive:2016xmw}, and we get the bare coupling $g=0.65$ by using the $O(\epsilon^2)$ correction in Eq.~(\ref{gCorrection}).

First, we list the results for the contact contributions. For $\mathcal{V}_{I=1}^{DD^*}$ in the channel of isospin 1, the effective potential at $O(\epsilon^0)$ and $O(\epsilon^2)$ is as follows
\begin{eqnarray}
\mathcal{V}^{(0)}_{I=1} &=& -2D_a+2D_b-2E_a+2E_b,  \label{PotentialV1} \\
\mathcal{V}^{(2)}_{I=1} &=& -(0.253+0.031i)D_b+0.044E_a \nonumber \\
 &&-(0.166+0.030i)E_b.\label{PotentialV3}
\end{eqnarray}
And for the channel of isospin 0, we obtain
\begin{eqnarray}
\mathcal{V}^{(0)}_{I=0} &=& -2D_a-2D_b+6E_a+6E_b,  \\
\mathcal{V}^{(2)}_{I=0} &=& -(1.214+0.190i)E_a+(0.116+0.047i)D_b \nonumber \\
&&+(0.025-0.143i)E_b.\label{PotentialV2}
\end{eqnarray}
Obviously, the contact contributions are just constants, and they result in $\delta(r)$ potentials in coordinate space, which describes short distance effect. From Eqs.~(\ref{PotentialV1})-(\ref{PotentialV2}),
we see the convergence of the series expansion is good. From Eqs.~(\ref{PotentialV3}) and (\ref{PotentialV2}), the contact coupling constant $D_a$ does not appear in the effective potential at $O(\epsilon^2)$ because the contributions from $D_a$ term are canceled among various diagrams in Fig.~\ref{O2ContactDiagram}.

\begin{figure}[htpb]
   	\begin{center}
   		\includegraphics[scale=0.3]{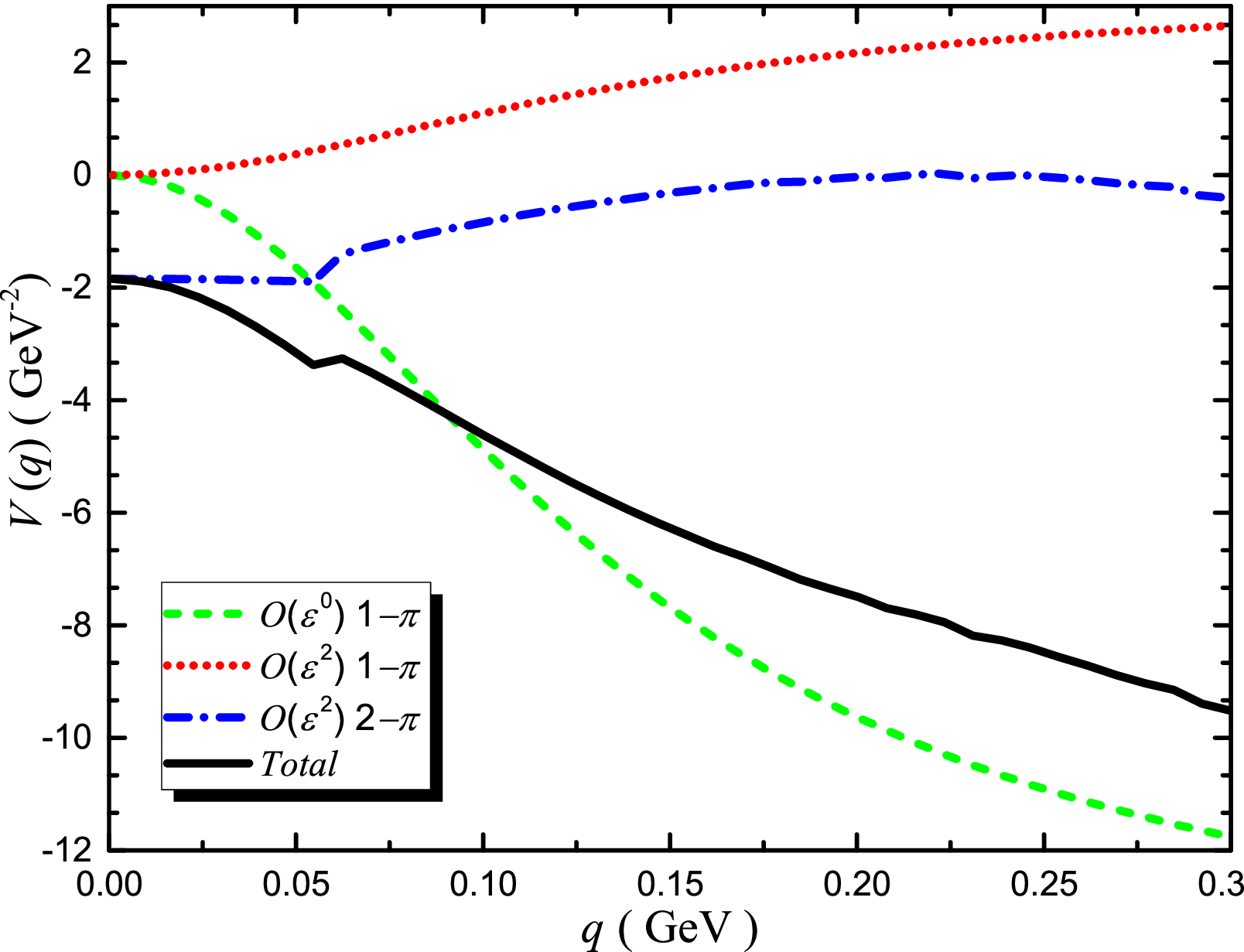}
   		\caption{(color online). OPE and TPE potentials $\mathcal{V}_{I=0}^{DD^*}$ for isospin-0 channel. $q$ stands for the three momentum in unit of GeV,  $y$ axis represents the effective potential in unit of GeV$^{-2}$. The red dotted and green dashed lines describe the OPE potentials at the leading and next to leading order, individually. The blue dot-dashed line is for the TPE potential. The sum of the three contributions is represented by the black solid line.
   			}\label{VMomentumCpDelta}
   	\end{center}
   \end{figure}

\begin{figure}[htpb]
   	\begin{center}
   		\includegraphics[scale=0.3]{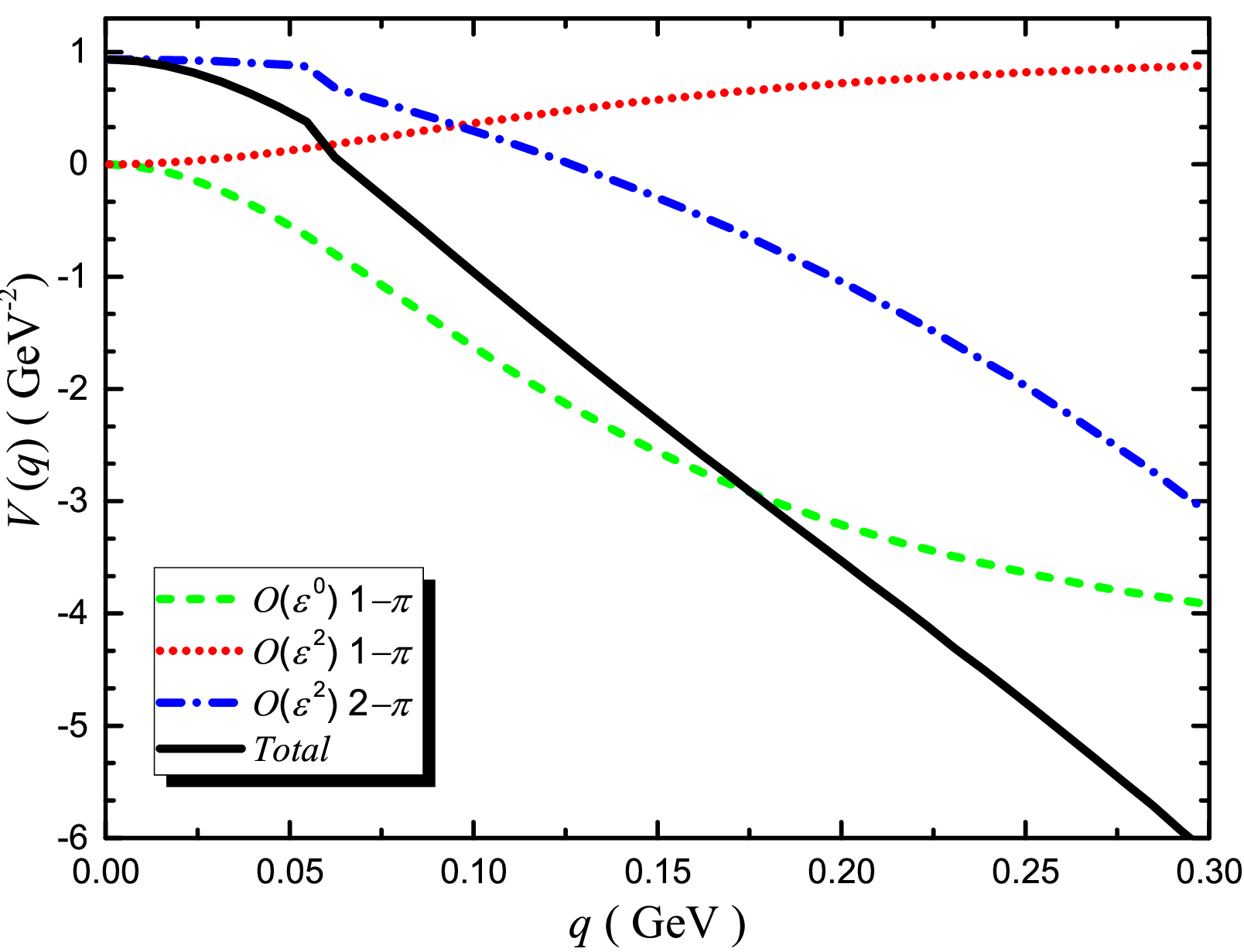}
   		\caption{(color online). OPE and TPE of the potentials $\mathcal{V}_{I=1}^{DD^*}$. The line types and color schemes match those of Fig. \ref{VMomentumCpDelta}.
   		  }\label{VMomentumCmDelta}
   	\end{center}
   \end{figure}

Next, we focus on the properties of the OPE and TPE contributions. We illustrate the corresponding potentials for channels with isospin 0 and 1
in Figs.~\ref{VMomentumCpDelta} and \ref{VMomentumCmDelta}, respectively, ranging from $q=|\mathbf{q}|=0$ to 300 MeV.

From Figs.~\ref{VMomentumCpDelta} and \ref{VMomentumCmDelta}, one can see that the OPE contributions at $O(\epsilon^0)$ are dominant in both the $I=0$ and $I=1$ channels since the green dashed lines are close to the black solid ones. The OPE potentials at $O(\epsilon^2)$
are small comparing to those at $O(\epsilon^0)$. The sums of OPE contributions are negative from Figs.~\ref{VMomentumCpDelta} and \ref{VMomentumCmDelta}, which means the OPE interaction is attractive in both the $I=0$ and $I=1$ channels. We also notice that the OPE interaction in $I=0$ is more attractive than $I=1$.

The situation for the TPE potentials is more complicated. The TPE contributions behave differently in $I=0$ and $I=1$ channels. In Fig.~\ref{VMomentumCpDelta}, the TPE interaction for $I=0$ channel is
attractive in the range $0\sim300$ MeV, and it tends to grow beyond $300$ MeV. The TPE potential at $O(\epsilon^2)$ is larger than the
OPE one at $O(\epsilon^0)$ in the range $0\sim60$ MeV, while the OPE contribution exceeds that of TPE rapidly when $q$ is larger than 60 MeV, and becomes dominant. We can say that the convergence of the chiral series is good. Looking at Fig.~\ref{VMomentumCmDelta}, we see the TPE potential is repulsive in the range $0 \sim 120$ MeV, while it becomes attractive as $q$ is beyond the range. The TPE potential at $O(\epsilon^2)$ is smaller than the OPE contribution at $O(\epsilon^0)$ in the lower range of the momentum, and becomes comparable in large
momenta. It seems to indicate the convergence of the chiral series would be spoiled at larger transfered momenta. From the blue dot-dashed lines in Figs. \ref{VMomentumCpDelta} and \ref{VMomentumCmDelta}, we see the TPE interaction in $I=1$ is more attractive than that in $I=0$.

Let us turn to the sum of these three contributions. The total contribution in Fig.~\ref{VMomentumCpDelta}
 for $\mathcal{V}_{I=0}^{DD^*}$ is attractive, while in Fig.~\ref{VMomentumCmDelta}
  for  $\mathcal{V}_{I=1}^{DD^*}$ it is less attractive and tends to repulsive as $q$ becomes smaller than 50 MeV because of the
  repulsive TPE contribution. It makes us wonder whether there could form a bound state in the $DD^*$ system with the inclusion of contact contributions.

\section{Potentials in coordinate space and possible molecular state} \label{secV}

Although the pion exchange interaction is attractive at most momenta, there can still be no bound states if not attractive enough. Moreover, the contact interaction might be repulsive and furthermore decrease the possibility for the existence of a bound state. Thus the contact potentials must be first obtained numerically by the determination of the LECs. After that, we can investigate the effective potentials in coordinate space, and then solve the Schr\"{o}dinger equation to search for possible molecular states.

\subsection{Determination of LECs}\label{secVA}
We determine the LECs in the contact contributions (\ref{PotentialV1})-(\ref{PotentialV2}) with the resonance saturation model \cite{Ecker:1988te,Donoghue:1988ed,Bernard:1996gq,Epelbaum:2001fm,Du:2016tgp}.
We assume these short-range couplings result from the $\rho$ and $\phi$ exchanges as in Ref.~\cite{Lu:2017dvm}, as well as
other meson exchanges (scalar and axial-vector). Although it may be a rough
estimate, it is meaningful to make such a attempt. The $D^{(*)}D^{(*)}V$ Lagrangian
respecting heavy quark symmetry and U(2) flavor symmetry is given by \cite{Li:2012ss}
\begin{eqnarray}\label{LagrangianHV}
\mathcal L_{HHV}&=&
i\beta\langle H v_\mu (V^\mu-\rho^\mu) \bar H \rangle
+i\lambda \langle H \sigma_{\mu\nu} F^{\mu\nu}(\rho) \bar H\rangle.
\end{eqnarray}
In the above, $H$ is the same as Eq.~(2), $F_{\mu\nu}=\partial_\mu\rho_\nu-\partial_\nu\rho_\mu-[\rho_\mu,\rho_\nu]$
with $\rho_\mu=\frac{ig_v}{\sqrt{2}}\hat{\rho}_\mu$, and the multiplet $\hat{\rho}$ is defined by
\begin{equation}
\hat{\rho}^\mu=\left(
\begin{array}{cc}
\frac{\rho^0}{\sqrt2}+\frac{\omega}{\sqrt2}&\rho^+\\
\rho^-&-\frac{\rho^0}{\sqrt2}+\frac{\omega}{\sqrt2}\\
\end{array}\right)^\mu.
\end{equation}
The coupling constants $g_v=5.8$, $\lambda=0.56$ GeV$^{-1}$ and $\beta=0.9$ \cite{Li:2012ss}. As for the scalar exchanges $S$~($\sigma$, $f_0$, $a_0$),
we use \cite{Li:2012ss,Du:2016tgp}:
\begin{eqnarray}\label{LagrangianHS}
\mathcal L_{HHS}&=&
g_{HHS}\langle H  S \bar H \rangle,
\end{eqnarray}
where $g_{HHa_0(f_0)}=\sqrt{3}g_{HH\sigma}$ \cite{Du:2016tgp}, $g_{HH\sigma}=\frac{g_\pi}{2\sqrt{6}}$ and $g_\pi=3.73$ \cite{Bardeen:2003kt}.
For the axial-vector mesons $A_V$~($a_1$, $f_1$), we use
\begin{eqnarray}\label{LagrangianHA}
\mathcal L_{HHA_V}&=&
g_{HHA_V}\langle H \gamma_\mu \gamma_5 A_V^\mu \bar H \rangle.
\end{eqnarray}
After matching the meson exchange amplitudes to the contact amplitudes with 4 independent isospin channels of $D^{(*)}D^{(*)}\to D^{(*)}D^{(*)}$, we obtain
\begin{eqnarray}\label{LECcontact}
D_a&=&-\frac{\beta^2g_v^2}{8m_\omega^2}-\frac{g_s^2}{2m_\sigma^2}-\frac{g_{s0}^2}{12m_{f_0}^2}, \quad E_a=-\frac{\beta^2g_v^2}{8m_\rho^2}-\frac{g_{s0}^2}{4m_{a_0}^2}, \nonumber\\
D_b&=& \frac{g_{HHA_V}^2}{8m_{a_1}^2}, \quad E_b=\frac{g_{HHA_V}^2}{8m_{f_1}^2}.
\end{eqnarray}
However, we can not find any inputs for axial-vector meson coupling $g_{HHA_V}$, and therefore we simply assume the low-energy constants are saturated by resonances with masses below 800 MeV. We estimate their errors with the contributions from the other four particles, $f_0$, $a_0$, $f_1$, and $a_1$. $|g_{HHA_V}|$ is roughly set to $\beta g_v\sim5$. We finally get the numerical values: 
\begin{eqnarray}\label{LECcontactValue}
D_a&=&-6.62\pm 0.15, \quad E_a=-5.74\pm 0.45, \nonumber\\
D_b&=& 0\pm 1.96, \quad E_b=0\pm 1.89.
\end{eqnarray}

\subsection{Potentials in coordinate space}
After the determination of the LECs, we are ready to transfer the potentials into coordinate space:
\begin{eqnarray}\label{FourierTransform}
\mathcal{V}(\mathbf{r})=\int \frac{d\mathbf{q}}{(2\pi)^3} \mathcal{V}(\mathbf{q})e^{i\mathbf{q} \cdot\mathbf{r}}.
\end{eqnarray}
However, since $\mathcal{V}(\mathbf{q})$ in ChEFT is proportional to the power series of $\mathbf{q}$, the higher order terms diverge worse. The evaluation of $\mathcal{V}(\mathbf{r})$ is essentially
a non-perturbative problem, and it originates from the resummation of the 2PI potentials. We have to regularize
Eq.~(\ref{FourierTransform}) non-perturbatively. Enormous efforts have been made to explore the non-perturbative
renormalization, such as Refs.~\cite{Ordonez:1992xp,Kaplan:1998tg,Epelbaum:1998ka,Nieves:2003uu,Yang:2004ss,PavonValderrama:2005wv,PavonValderrama:2005uj,Valderrama:2016koj}.
Here we resort to a simple Gaussian cutoff exp$(-\vec{p}^{2n}/\Lambda^{2n})$ to suppress the higher momentum
contributions, as in Ref.~\cite{Ordonez:1995rz,Epelbaum:1999dj,Ren:2016jna}. We use $n=2$ as in Ref.~\cite{Ren:2016jna}.
In the nucleon-nucleon ChEFT, the value of cutoff parameter is commonly below the $\rho$ meson
 mass \cite{Epelbaum:2014efa}, and therefore we adopt $\Lambda=0.7$ GeV in our work.

\begin{figure}[htpb]
   	\begin{center}
   		\includegraphics[scale=0.3]{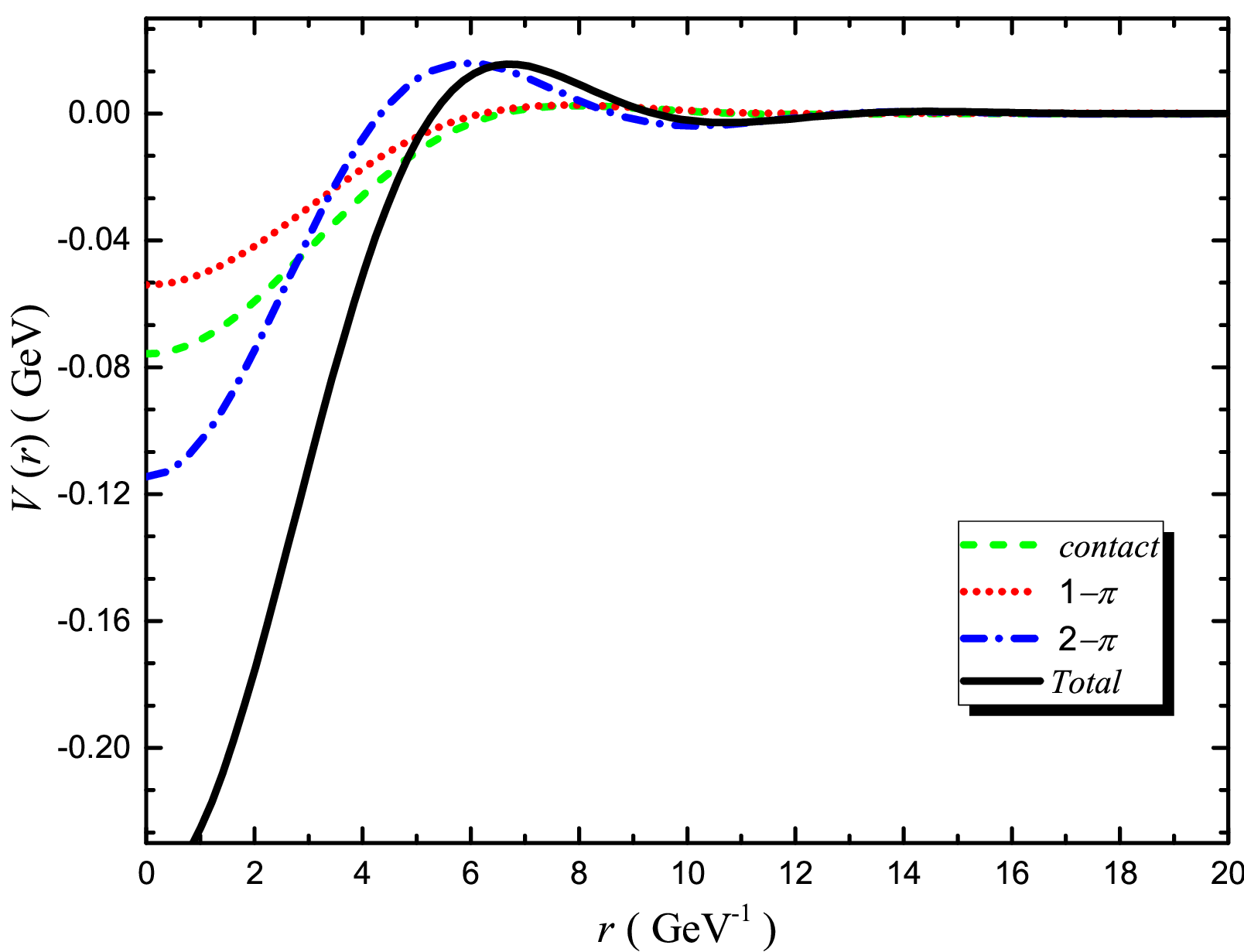}
   		\caption{(color online). $S$-wave potentials of the $DD^*$ systerm with $I=0$ in units of GeV. The green dashed, red dotted, and blue dot-dashed lines stand for the contact, OPE, and TPE contributions, respectively. The full potential is drawn in black solid line.
   		}\label{Vr}
   	\end{center}
   	   \end{figure}

\begin{figure}[htpb]
   	      	   	\begin{center}
   	      	   		\includegraphics[scale=0.3]{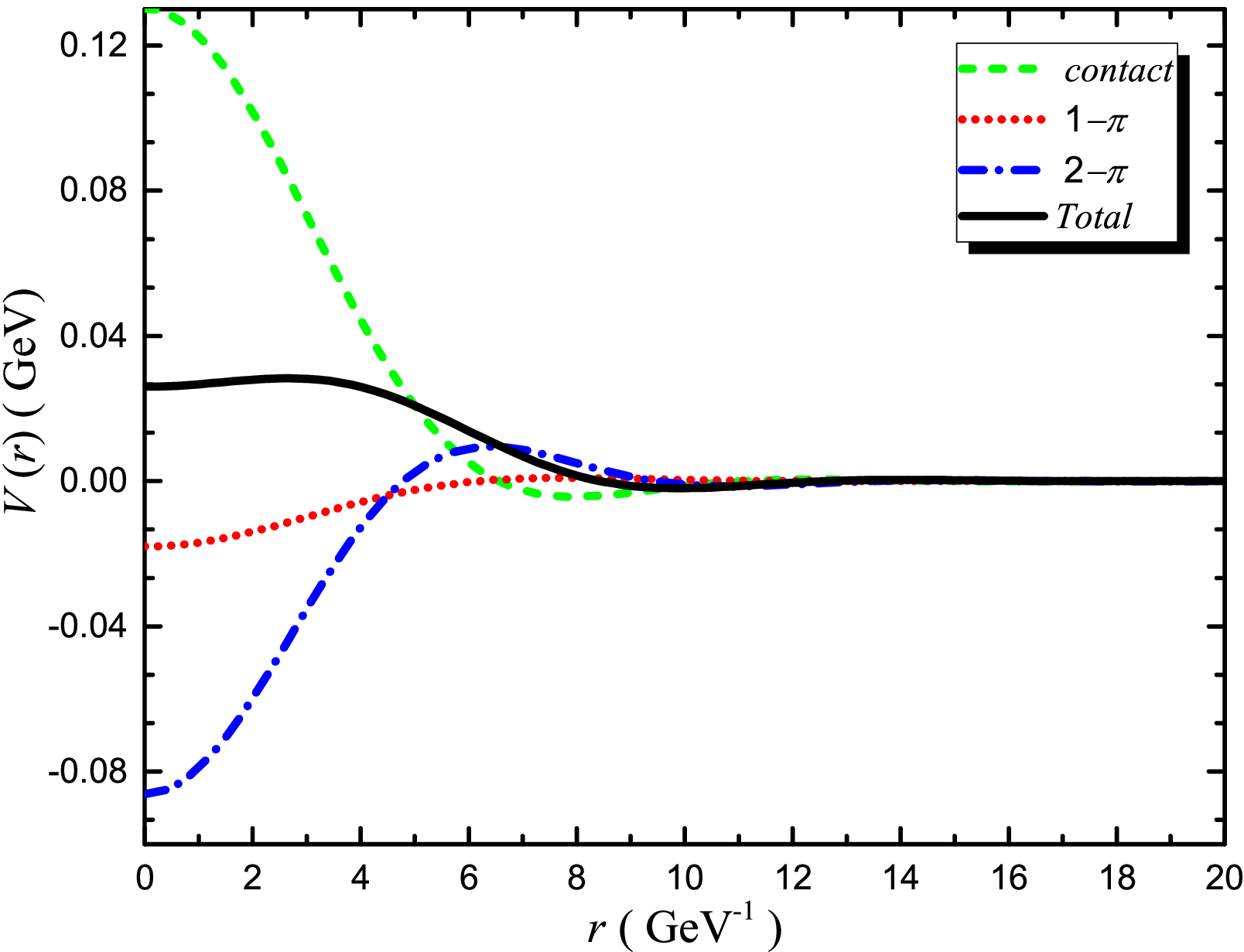}
   	      	   		\caption{(color online). $S$-wave potentials of the $DD^*$ systerm with $I=1$ in units of GeV. The line types and color schemes match those of Fig. \ref{Vr}.
   	      	   		}\label{Vrs}
   	      	   	\end{center}
   	      \end{figure}

The resulting full potentials are shown in Figs.~\ref{Vr} and \ref{Vrs}, where we set $\Lambda=0.7$ GeV. From Figs.~\ref{Vr} and \ref{Vrs}, we find the OPE and TPE interactions are attractive in both cases, and the contact terms lead to the attractive interaction in the $I=0$ channel while repulsive interaction in the $I=1$ channel.
 Obviously, this difference brings more opportunity to form a bound state in the $I=0$ channel than in the $I=1$ channel.
 Let us focus on the total results. The
 total potential in the short distance for the $I=1$ channel is repulsive but small while that for the $I=0$ channel is attractive and large.
      \begin{figure}[htpb]
      	\begin{center}
      		\includegraphics[scale=0.3]{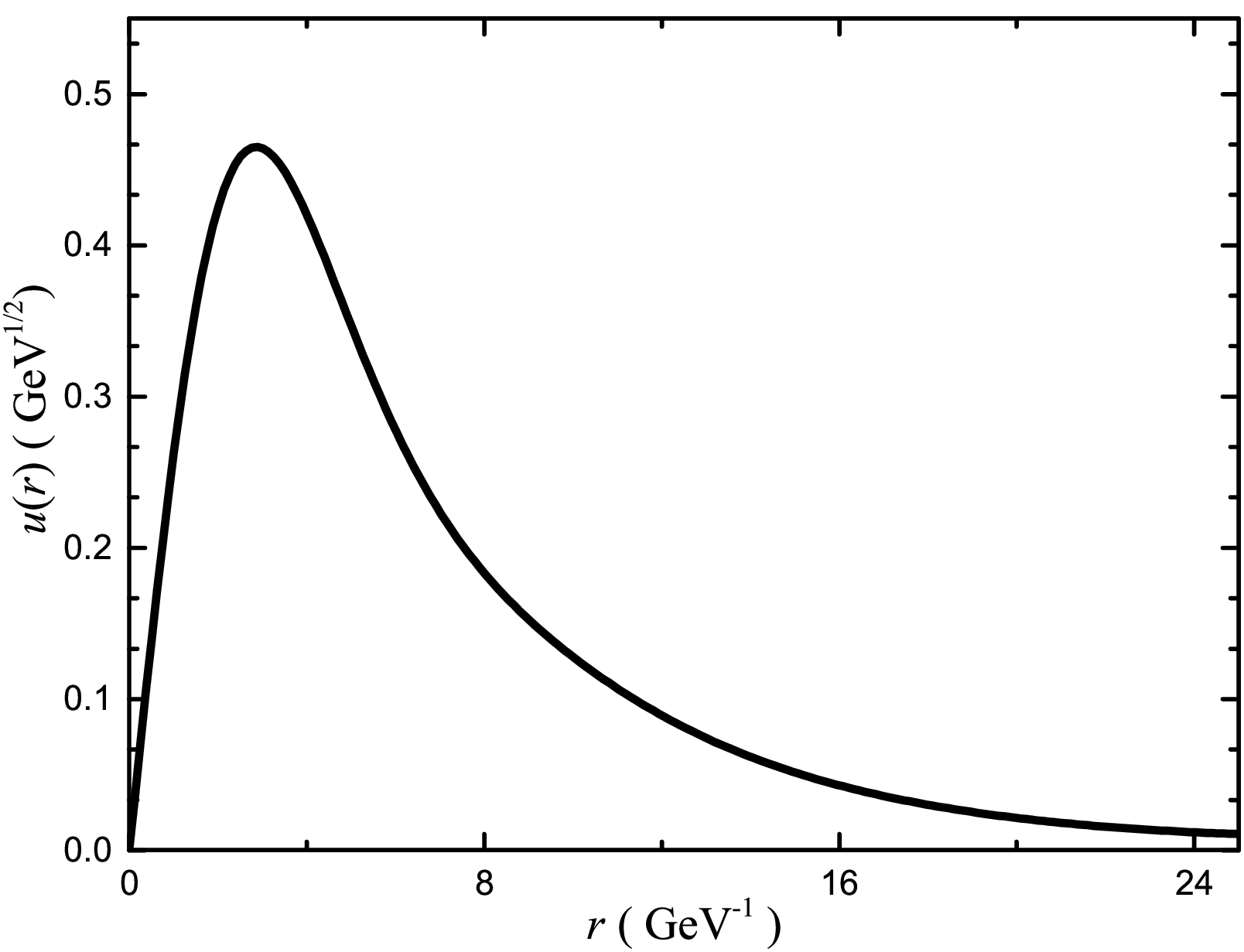}
      		\caption{(color online). The radial wave function with the full potential depicted in Fig.~\ref{Vr}.
      		}\label{Psir}
      	\end{center}
      \end{figure}

\subsection{Possible bound states}
 With the potentials in hand, we are finally able to solve Schr\"{o}dinger equation. We find a bound state with the binding energy around $17.5$ MeV in the $I=0$ channel, and there exists no bound state in the $I=1$ channel.

The radial wave function for the $I=0$ channel is plotted in Fig.~\ref{Psir}. It extends to quite large distance, which means the constituents $D$ and $D^*$ are separated.

It is worth noticing that in pion and vector-meson exchange potential model \cite{Ohkoda:2012hv} a bound state was found
with a binding energy of $62.3$ MeV in $I=0$ channel, while no state was found in $I=1$ channel.  In the one-boson exchange model, there is also a bound state in the $I=0$ channel, and the binding energy is about $5\sim43$ MeV with a reasonable cutoff \cite{Li:2012ss}.  No bound
state was found in the $I=1$ channel in that model, either \cite{Li:2012ss}. Our results are consistent.

{
From Fig. \ref{Vrs}, we notice the contact interaction is repulsive at short distance. However, we cannot still find a bound state even if dropping the contact interaction in the $I=1$ channel, which states the pion exchange interaction is not attractive enough for binding $DD^*$. If we repeat and turn off the contact potential in the $I=0$ channel, the shallow bound state will disappear. We cannot obtain a reasonable energy eigenvalue of the Schr\"odinger equations, either, if keeping the OPE potentials themselves for two channels. The attractive contact and TPE interactions are important for the existence of the molecule in the $I=0$ channel.
}

\begin{figure}[htpb]
    	   	\begin{center}
    	   		\includegraphics[scale=0.3]{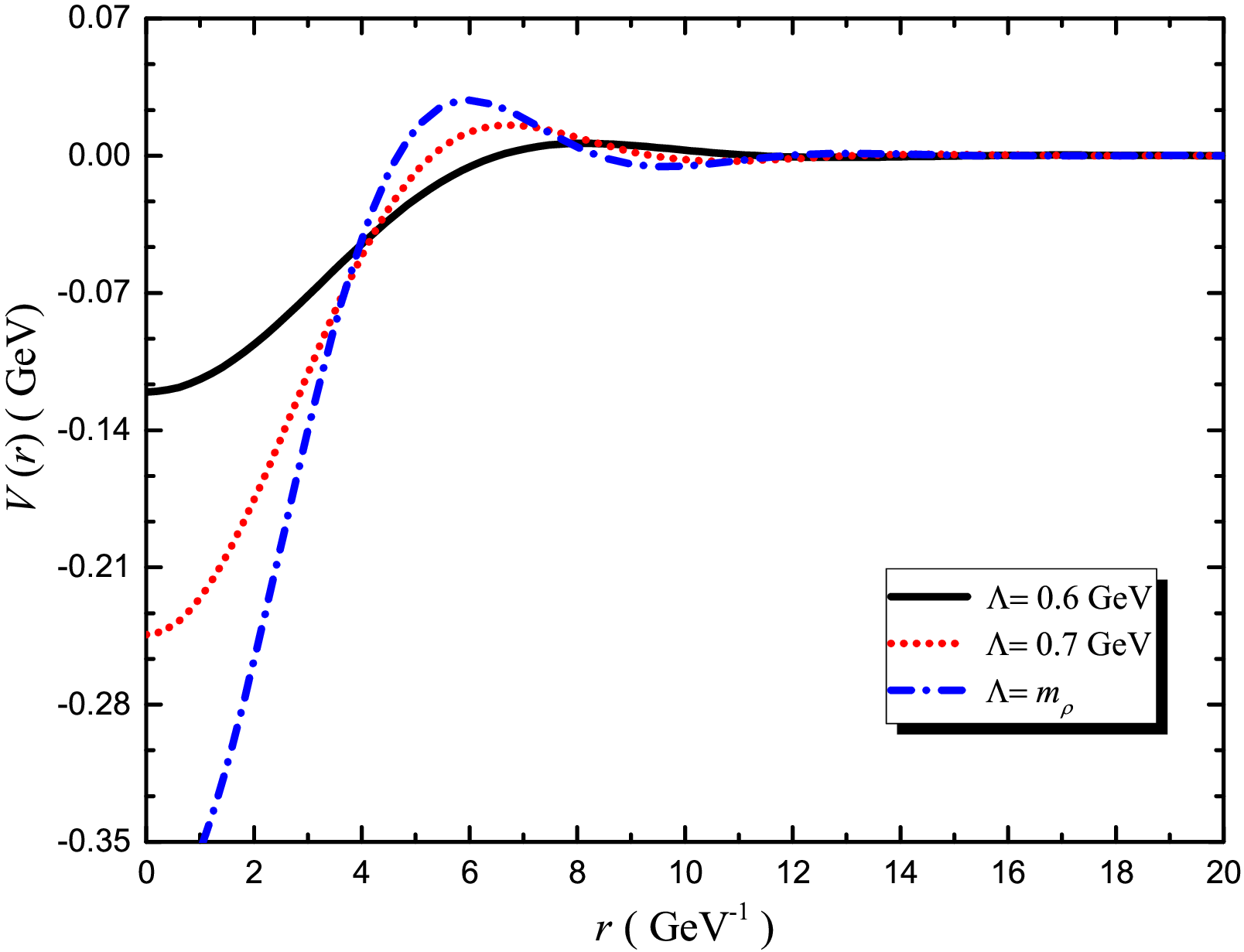}
    	   		\caption{(color online). The total potentials of the $DD^*$ system in the $S$ wave with $I=0$, where three cutoff values are adopted.
    	   		}\label{VrpsL}
    	   	\end{center}
    	   \end{figure}

Theoretically, the obtained observable (such as binding energy) is independent of the regularization procedure in Eq.~(\ref{FourierTransform}). The formal dependence on the cutoff $\Lambda$ in Eq.~(\ref{FourierTransform}) can be compensated by the $\Lambda$ dependence of the LECs. However, the results are sometimes sensitive with different choices of $\Lambda$ in practice. Here we investigate the influence of the cutoff with the LECs fixed. We plot the full potentials with different cutoffs in Fig.~\ref{VrpsL}. From the figure, we notice that the potential becomes deeper and steeper in the short range as the cutoff increases. After solving Schr\"{o}dinger equation, we obtain the binding energy $1.1$ MeV, $17.5$ MeV and $53.1$ MeV with $\Lambda=0.6$, $0.7$ GeV, and $m_\rho$, respectively. The binding energy is sensitive to the cutoff. However, bound state solution exists as cutoff is near $m_\rho$. Furthermore, as we stressed earlier, the cutoff dependence can be compensated if readjusting the LECs at different cutoffs.

There also exist other sources of uncertainties. Firstly, we discuss the uncertainty from the resonance saturation model which is utilized to determine LECs of contact terms. From the numerical
values of $D_a$ and $E_a$ in Eq. \ref{LECcontactValue}, we can see the contributions from $f_0$ and $a_0$ are small, and $\rho$, $\omega$
and $\sigma$ exchanges dominate $D_a$ and $E_a$. For $D_b$ and $E_b$ in Eq. \ref{LECcontactValue}, the uncertainty brought by
axial-vector exchanges are not small, and therefore they have considerable effects on binding energy. However 
the estimation of the axial-vector contributions is quite rough, we hope we can obtain much more reliable input for $g_{HHV_A}$ in the future. In general, the uncertainty in Eq~(\ref{LECcontactValue}) gives the binding energy at $\Lambda=0.7$ GeV: $17.5^{+4.1+18.3}_{-3.9-14.0}$ MeV, where the first uncertainty comes from $f_0$ and $a_0$, 
and the second uncertainty comes from axial-vector mesons ($a_1$, $f_1$).

Secondly, the uncertainty can come from that of axial coupling $g$. When we include the experimental error \cite{Olive:2016xmw} (width and branching fraction), we obtain the bare coupling $g=0.65^{+0.02}_{-0.01}$, and the binding energy in $I=0$ channel with $\Lambda=0.7$ GeV is $17.5^{+9.6}_{-3.9}$ MeV. We can see the binding energy is sensitive to the coupling $g$, but not much sensitive as cutoff
$\Lambda$. Including the uncertainty of $D_{a(b)}$, $E_{a(b)}$ discussed above, we obtain the binding energy $17.5^{+21.1}_{-15.0}$ at $\Lambda=0.7$
GeV. This uncertainty is largely brought by axial-vector mesons, the uncertainty from $g$ is moderate, and the uncertainty from $f_0$ and $a_0$ is smallest.

\begin{figure}[htpb]
	\begin{center}
		\includegraphics[scale=0.45]{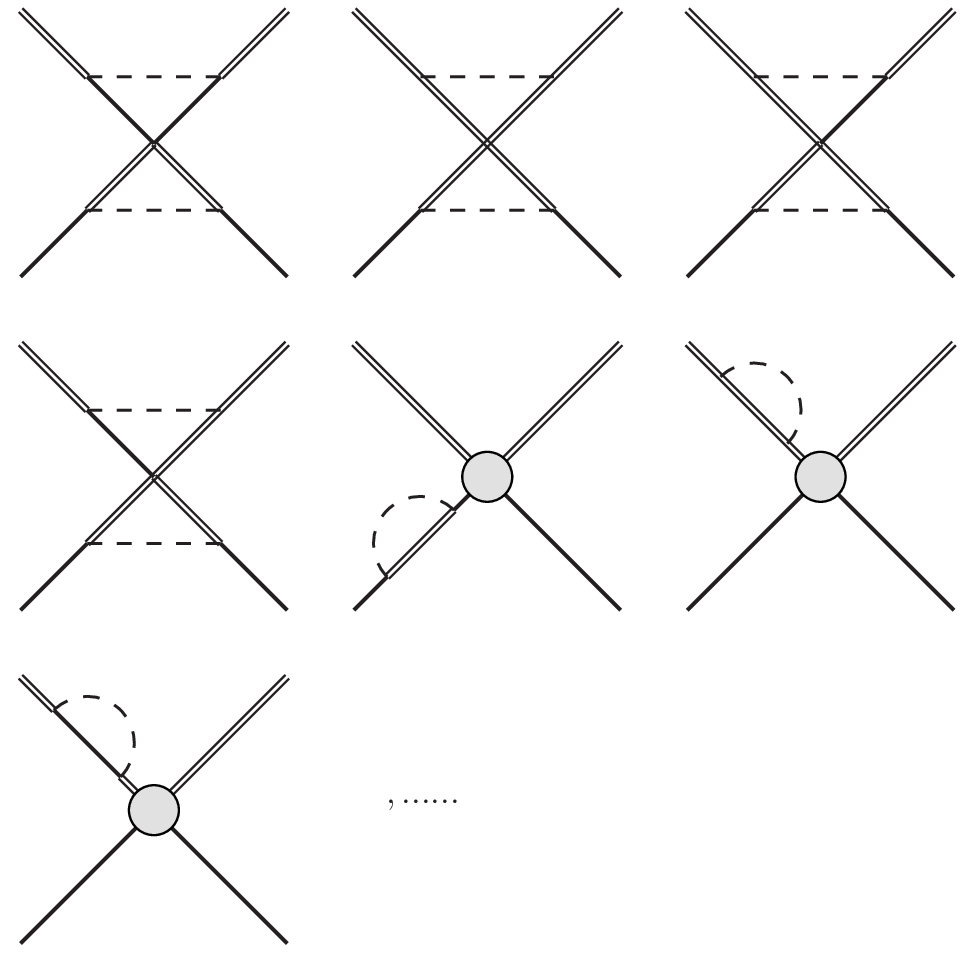}
		\caption{Some loop diagrams related to contact terms at $O(\epsilon^4)$. The last three diagrams indicate the contribution from wavefucntion renormalization to the contact loop diagrams at $O(\epsilon^2)$ in Fig. 3.
		}\label{O4ContactDiagrams}
	\end{center}
\end{figure}

The third uncertainty comes from truncation error. Here we partially estimate few loop diagrams of contact contribution at $O(\epsilon^4)$ to show how large the truncation error is. For $O(\epsilon^4)$ contact loop contribution, there exist many Feynmann diagrams. We pick some diagrams and plot them in Fig.~\ref{O4ContactDiagrams}.
In the first four diagrams of Fig.~\ref{O4ContactDiagrams}, each one contains two separated loops, and the sum of these reads:
\begin{eqnarray}
\mathcal{V}^{(4)}_{I=1} &\sim& (-0.00016-0.00015i)(D_a-D_b+E_a-E_b), \nonumber \\
\mathcal{V}^{(4)}_{I=0} &\sim& (-0.0039-0.0014i)D_a+(0.0117+0.0041i)E_a \nonumber \\
&&+(0.0043+0.0019i)D_b-(0.0129+0.0057i)E_b. \nonumber\\
\end{eqnarray}	
We can see they are generally $O(\frac{1}{100})$ relative to those at $O(\epsilon^2)$ by comparing with Eqs. (21) and (23).
The last three set of diagrams in Fig.~\ref{O4ContactDiagrams} indicate the wave function renormalization of the $O(\epsilon^2)$ diagrams
in Fig.~3, and the sum of these reads:
\begin{eqnarray}
\mathcal{V}^{(4)}_{I=1} &\sim& (0.0196+0.0060i)D_b-0.0034E_a \nonumber \\
&&+(0.0127+0.0047i)E_b,\\
\mathcal{V}^{(4)}_{I=0} &\sim& (0.0934+0.0321i)E_a-(0.0085+0.0054i)D_b \nonumber \\
&&-(0.0039-0.0110i)E_b.
\end{eqnarray}
They are $O(\frac{1}{10})$ relative to those at $O(\epsilon^2)$ from Eqs.~ (21) and (23). Therefore we expect
when all the contact $O(\epsilon^4)$ diagrams are included, the convergence may not be bad.

\section{summary}

In this work, we have systematically studied the $DD^*$ system with ChEFT. Due to the intrinsic difficulty of the ChEFT, we cannot obtain the physical observables directly from the Feynman diagrams. We alternatively calculate the potentials, i.e., the sum of all the 2PI diagrams, and then iterate them into Lippmann-Schwinger or Schr\"odinger equation to recover the 2PR contributions.

We have investigated the $DD^*$ effective potentials in ChEFT with Weinberg scheme. With the effective potentials obtained in momentum space, we have analyzed the contact, OPE and TPE contribution in detail. The OPE and TPE contributions are free of many LECs, and thus they are more model independent than the contact interaction since the LECs are determined with the resonance saturation model in this work. The OPE contribution at $O(\epsilon^2)$ is smaller than that at $O(\epsilon^0)$. The potential from TPE at $O(\epsilon^2)$ is relatively large compared to that from OPE at $O(\epsilon^0)$ in the $I=1$ channel, while it shows a good convergence in the $I=0$ channel. The TPE interaction is important and non-negligible.

We have determined the LECs in contact contributions with the resonance saturation model, and further explored the full potentials in coordinate space, which are regularized with a simple Gaussian cutoff. The roles of each contributions have been discussed, and the total potentials are very different in two channels. We have also discussed the importance of the contact contribution and the infulence of the cutoff in detail. Furthermore, we discuss the uncertainties of our approach, which comes from axial coupling $g$, LECs and truncation error. We find that the TPE contribution is non-negligible and attractive in general, while the contact contributions is an important element to compete the $\pi$-exchange contributions and cause quite different behavior in each channels. Despite the roughly estimated LECs, we notice that no bound state exists in the $I=1$ channel in a wide range of cutoff parameter, while there is a bound state in the $I=0$ channel as the cutoff is near $m_\rho$ in our approach. The binding energy is sensitive to the cutoff. Our results are consistent with those in the one-boson-exchange model \cite{Li:2012ss}.

In this work, we have ignored many other sub-leading effects from the isospin violation, $S$-$D$ mixing, recoiling, and so on. These effects can be investigated in future, and our framework shall be proved to be elegant.

We point out that the $DD^*$ molecule may be discovered at experiments through various processes. Since at Tevatron and LHCb there are a number of $B_c$ events, the $DD^*$ molecule can be produced via $B_c$ weak decay: singly Cabibbo-supressed process $B_c\to X(DD^*) K$, doubly Cabibbo-supressed processes $B_c\to X(DD^*) \pi$ and $B_c\to X(DD^*) D$. Moreover we hope the $e^+e^-$ process such
as $e^+e^-\to X(DD^*) \bar{D}\bar{D}$ at BelleII can be studied to observe the state. The molecular states may be constructed through $DD$ final states. We also expect the lattice simulations to test our results.

Our exploration of the $DD^{*}$ system can help to make more profound understanding of the heavy meson system and non-perturbative QCD. We  expect our results can be tested by future LHCb and BelleII experiments, and help the extrapolations of future lattice simulations.

\section*{Acknowledgments}
We would like to thank Professor Shi-Lin Zhu for useful suggestion. Xu also thanks Rui Chen and Ming-Xiao Duan for helpful discussions. This project is supported by the National Natural Science Foundation of China under Grants No. 11705072. This work is also supported by the Fundamental Research Funds for the Central Universities. Xiang Liu is also supported by the China National Funds for Distinguished Young Scientists under Grants No. 11825503.

\appendix
\section{One-loop amplitudes of the processes
	 $DD^{*}\to DD^{*}$ at $O(\epsilon^2)$} \label{SecAppA}
We first list the amplitudes of the process $D(p_1)D^{*}(p_2)\to D(p_3)D^{*}(p_4)$. The difference between the amplitudes for the $I=0$ and $I=1$ channels is just a factor.

For the one-loop corrections to the contact terms in Fig.~\ref{O2ContactDiagram}, the Feynman amplitudes are
\begin{eqnarray}
&&\mathcal{M}^{(2)}_{(a1)} = \frac{-i}{4} \frac{g^2}{f^2} A_{a1} J^g_{22}(m,\omega_1,\omega_2) \varepsilon(p_2)\cdot\varepsilon^*(p_4)  \nonumber \\  \label{AllAmplitude1}
 && \textnormal{with}  \;    \omega_1 = v \cdot p_2 - M, \; \omega_2 = v \cdot p_4 - M;  \\
&&\mathcal{M}^{(2)}_{(a2)} = \frac{-i}{2} \frac{g^2}{f^2} A_{a2} J^g_{22}(m,\omega_1,\omega_2)  \varepsilon(p_2)\cdot\varepsilon^*(p_4) \nonumber\\
 && \textnormal{with}  \;    \omega_1 = v \cdot p_2 - M-\delta, \; \omega_2 = v \cdot p_4 - M-\delta;        \\
&&\mathcal{M}^{(2)}_{(a3)} = \frac{i}{4} \frac{g^2}{f^2} A_{a3} J^g_{22}(m,\omega_1,\omega_2) \varepsilon(p_2)\cdot\varepsilon^*(p_4)  \nonumber\\
 && \textnormal{with}  \;    \omega_1 = v \cdot p_2 - M, \; \omega_2 = v \cdot p_3 - M-\delta;        \\
&&\mathcal{M}^{(2)}_{(a4)} = \frac{-i}{4} \frac{g^2}{f^2} A_{a4} J^g_{22}(m,\omega_1,\omega_2) \varepsilon(p_2)\cdot\varepsilon^*(p_4) \nonumber\\
 && \textnormal{with}  \;    \omega_1 = v \cdot p_1 - M-\delta, \; \omega_2 = v \cdot p_3 - M-\delta;        \\
&&\mathcal{M}^{(2)}_{(a5)} = \frac{i}{4} \frac{g^2}{f^2} A_{a5} J^g_{22}(m,\omega_1,\omega_2) \varepsilon(p_2)\cdot\varepsilon^*(p_4) \nonumber\\
 && \textnormal{with}  \;    \omega_1 = v \cdot p_1 - M-\delta, \; \omega_2 = v \cdot p_4 - M;        \\
&&\mathcal{M}^{(2)}_{(a6)} = \frac{i}{4} \frac{g^2}{f^2} A_{a6} J^h_{22}(m,\omega_2,\omega_1) \varepsilon(p_2)\cdot\varepsilon^*(p_4)  \nonumber \\
 && \textnormal{with}  \;    \omega_1 = v \cdot p_1 - M-\delta, \; \omega_2 = v \cdot p_2 - M;  \\
&&\mathcal{M}^{(2)}_{(a7)} = \frac{i}{4} \frac{g^2}{f^2} A_{a7} J^h_{22}(m,\omega_2,\omega_1) \varepsilon(p_2)\cdot\varepsilon^*(p_4)  \nonumber \\
 && \textnormal{with}  \;    \omega_1 = v \cdot p_4 - M, \; \omega_2 = v \cdot p_3 - M-\delta;  \\
&&\mathcal{M}^{(2)}_{(a8)} = \frac{-i}{2} \frac{g^2}{f^2} A_{a8} J^h_{22}(m,\omega_2,\omega_1) \varepsilon(p_2)\cdot\varepsilon^*(p_4)  \nonumber \\
 && \textnormal{with}  \;    \omega_1 = v \cdot p_1 - M-\delta, \; \omega_2 = v \cdot p_2 - M-\delta;  \\
&&\mathcal{M}^{(2)}_{(a9)} = \frac{-i}{2} \frac{g^2}{f^2} A_{a9} J^h_{22}(m,\omega_2,\omega_1) \varepsilon(p_2)\cdot\varepsilon^*(p_4)  \nonumber \\
 && \textnormal{with}  \;    \omega_1 = v \cdot p_4 - M-\delta, \; \omega_2 = v \cdot p_3 - M-\delta;  \\
&&\mathcal{M}^{(2)}_{(a10)} = \frac{i}{2} \frac{g^2}{f^2} A_{a10} J^g_{22}(m,\omega_1,\omega_2) \varepsilon(p_2)\cdot\varepsilon^*(p_4)  \nonumber \\
 && \textnormal{with}  \;    \omega_1 = v \cdot p_2 - M-\delta, \; \omega_2 = v \cdot p_3 - M-\delta;  \\
&&\mathcal{M}^{(2)}_{(a11)} = \frac{i}{2} \frac{g^2}{f^2} A_{a11} J^g_{22}(m,\omega_1,\omega_2) \varepsilon(p_2)\cdot\varepsilon^*(p_4)  \nonumber \\
 && \textnormal{with}  \;    \omega_1 = v \cdot p_1 - M-\delta, \; \omega_2 = v \cdot p_4 - M-\delta;  \\
&&\mathcal{M}^{(2)}_{(a12+a13)} \nonumber \\
 &&= -i \frac{g^2}{f^2} A_{a12a13}  \bigg( \frac{3}{8}\partial\omega J^b_{22}(m,\omega_1)  + \frac{3}{4}\partial\omega J^b_{22}(m,\omega_2) \bigg)  \nonumber \\
 && \quad \times \varepsilon(p_2)\cdot\varepsilon^*(p_4)  \nonumber\\
 && \textnormal{with}  \;    \omega_1 = v \cdot p_2 - M, \; \omega_2 = v \cdot p_2 - M-\delta,  \textnormal{ and with} \nonumber     \\
 &&  \omega_1 = v \cdot p_4 - M, \; \omega_2 = v \cdot p_4 - M-\delta;  \\
&&\mathcal{M}^{(2)}_{(a14)} \nonumber \\
 &&= -i \frac{g^2}{f^2} A_{a14} \bigg( \frac{9}{8} \partial\omega J^b_{22}(m,\omega_1) \bigg)    \varepsilon(p_2)\cdot\varepsilon^*(p_4)  \nonumber \\
 && \textnormal{with}  \;    \omega_1 = v \cdot p_1 - M-\delta,   \textnormal{ and with } \omega_1 = v \cdot p_3 - M-\delta. \nonumber\\
\end{eqnarray}

For the one-loop corrections to the OPE potentials in Fig.~\ref{O21piDiagram}, the Feynman amplitudes are
\begin{eqnarray}
&&\mathcal{M}^{(2)}_{(b1)} = \frac{i}{4}\frac{g^4}{f^4}A_{b1} \frac{p_\mu p_\nu}{p^2-m^2} J^g_{22}(m,\omega_1,\omega_2) \varepsilon^\mu(p_2)\varepsilon^{*\nu}(p_4) \nonumber \\
&& \textnormal{with}  \;    \omega_1 = v \cdot p_3 - M-\delta, \;\omega_2 = v \cdot p_2 - M; \\
&&\mathcal{M}^{(2)}_{(b2)} = \frac{i}{2}\frac{g^4}{f^4} A_{b2}\frac{p_\mu p_\nu}{p^2-m^2} J^g_{22}(m,\omega_1,\omega_2) \varepsilon^\mu(p_2)\varepsilon^{*\nu}(p_4) \nonumber \\
&& \textnormal{with}  \;    \omega_1 = v \cdot p_2 - M-\delta, \;\omega_2 = v \cdot p_3 - M-\delta; \\
&&\mathcal{M}^{(2)}_{(b3)} = \frac{i}{4}\frac{g^4}{f^4} A_{b3} \frac{p_\mu p_\nu}{p^2-m^2} J^g_{22}(m,\omega_1,\omega_2) \varepsilon^\mu(p_2)\varepsilon^{*\nu}(p_4) \nonumber \\
&& \textnormal{with}  \;    \omega_1 = v \cdot p_1 - M-\delta, \;\omega_2 = v \cdot p_4 - M; \\
&&\mathcal{M}^{(2)}_{(b4)} = \frac{i}{2}\frac{g^4}{f^4} A_{b4} \frac{p_\mu p_\nu}{p^2-m^2} J^g_{22}(m,\omega_1,\omega_2) \varepsilon^\mu(p_2)\varepsilon^{*\nu}(p_4) \nonumber \\
&& \textnormal{with}  \;    \omega_1 = v \cdot p_1 - M-\delta, \;\omega_2 = v \cdot p_4 - M-\delta; \\
&&\mathcal{M}^{(2)}_{(b5)} = i \frac{g^2}{f^4} A_{b5} \frac{p_\mu p_\nu}{p^2-m^2} \bigg[2m^2L+\frac{2m^2}{16\pi^2}\textnormal{log}(\frac{m}{\mu}) \bigg] \nonumber \\
&&  \qquad \quad \ \  \times \varepsilon^\mu(p_2)\varepsilon^{*\nu}(p_4);  \\
&&\mathcal{M}^{(2)}_{(b6)} = i\frac{g^2}{f^4} A_{b6} \frac{p_\mu p_\nu}{p^2-m^2} \bigg[2m^2L+\frac{2m^2}{16\pi^2}\textnormal{log}(\frac{m}{\mu}) \bigg] \nonumber \\
&&  \qquad \quad \ \  \times \varepsilon^\mu(p_2)\varepsilon^{*\nu}(p_4);  \\
&&\mathcal{M}^{(2)}_{(b7)} = i\frac{g^2}{f^2} A_{b7} \frac{p_\mu p_\nu}{p^2-m^2} \bigg[\frac{2}{3f^2}\Big(2m^2L+\frac{2m^2}{16\pi^2}\textnormal{log}(\frac{m}{\mu})\Big) \bigg] \nonumber \\
&&  \qquad \quad \ \  \times \varepsilon^\mu(p_2)\varepsilon^{*\nu}(p_4);  \\
&&\mathcal{M}^{(2)}_{(b8)} =0; \quad \mathcal{M}^{(2)}_{(b9)} =0; \\
&&\mathcal{M}^{(2)}_{(b10+b11)}  \nonumber \\
&&= -i\frac{g^4}{f^4} A_{b10b11} \frac{p_\mu p_\nu}{p^2-m^2} \bigg(\frac{3}{8}\partial\omega J^b_{22}(m,\omega_1)  +\frac{3}{4}\partial\omega J^b_{22}(m,\omega_2)\bigg) \nonumber \\
&& \quad \times \varepsilon^\mu(p_2)\varepsilon^{*\nu}(p_4) \nonumber \\
&& \textnormal{with}  \;    \omega_1 = v \cdot p_2 - M, \;\omega_2 = v \cdot p_2 - M-\delta, \textnormal{ and with} \nonumber \\
&& \omega_1 = v \cdot p_4 - M, \;\omega_2 = v \cdot p_4 - M-\delta ;\\
&&\mathcal{M}^{(2)}_{(b12)}  \nonumber \\
&&= -i\frac{g^4}{f^4} A_{b12} \frac{p_\mu p_\nu}{p^2-m^2} \bigg(\frac{9}{8}\partial\omega J^b_{22}(m,\omega_1) \bigg) \varepsilon^\mu(p_2)\varepsilon^{*\nu}(p_4) \nonumber \\
&& \textnormal{with}  \;    \omega_1 = v \cdot p_1 - M-\delta, \textnormal{ and with } \omega_1 = v \cdot p_3 - M-\delta;  \nonumber  \\
\end{eqnarray}

For the TPE potentials in Fig.~\ref{O22piDiagram}, the Feynman amplitudes are
\begin{eqnarray}
&&\mathcal{M}^{(2)}_{(c1)} \nonumber\\
 &&= \frac{i}{4f^4} \bigg[ 4 A_{c1a}\Big(q^2_0J^F_{21}+J^F_{22}\Big) + 4 A_{c1b}q^2_0J^F_{11} +A_{c1c} J^F_0 \bigg]  \nonumber \\
 && \quad \times \varepsilon(p_2)\cdot\varepsilon^*(p_4); \\
&&\mathcal{M}^{(2)}_{(c2)} \nonumber\\
 &&= \frac{-i}{4}\frac{g^2}{f^4} A_{c2} \bigg[ \Big(2 A_{c2c} q_0J^S_{31}+2A_{c2c}J^S_{34}+A_{c2d}q_0J^S_{21}\Big) \nonumber \\
 && \quad \times  \varepsilon(p_2)\cdot\varepsilon^*(p_4) + \Big(2A_{c2c} q_0J^S_{32} +2A_{c2c} J^S_{33} \nonumber \\
 && \quad +(A_{c2d}+2A_{c2c})q_0J^S_{22}+2A_{c2c}J^S_{24}+A_{c2d}q_0J^S_{11}\Big) \nonumber \\
 && \quad \times q\cdot\varepsilon(p_2) q\cdot\varepsilon^*(p_4) \bigg] \nonumber \\
 && \textnormal{with}  \;    \omega = v \cdot p_2 - M;  \\
&&\mathcal{M}^{(2)}_{(c3)} \nonumber\\
 &&= \frac{i}{4}\frac{g^2}{f^4} A_{c3}\bigg[ \Big(2A_{c3d}q_0J^S_{21}-(2A_{c3c}+A_{c3d})q_0\vec{q}^2J^S_{22} \nonumber \\
 && \quad -2A_{c3c}\vec{q}^2J^S_{24}+4A_{c3c}q_0J^S_{31}-2A_{c3c}q_0\vec{q}^2J^S_{32}-2A_{c3c}\vec{q}^2J^S_{33} \nonumber\\
 && \quad +4A_{c3c}J_{34}-A_{c3d}q_0\vec{q}^2J^S_{11}\Big) \varepsilon(p_2)\cdot\varepsilon^*(p_4) + \Big( -A_{c3d}q_0J^S_{11} \nonumber \\
 && \quad -(2A_{c3c}+A_{c3d})q_0J^S_{22}-2 A_{c3c}J^S_{24}-2 A_{c3c}q_0J^S_{32} \nonumber \\
 && \quad -2A_{c3c}J^S_{33} \Big) q\cdot\varepsilon(p_2) q\cdot\varepsilon^*(p_4) \bigg] \nonumber\\
 && \textnormal{with}  \;    \omega = v \cdot p_2 - M-\delta ; \\
&&\mathcal{M}^{(2)}_{(c4)} \nonumber\\
 &&= \frac{i}{4}\frac{g^2}{f^4} \bigg[ A_{c4d}q_0\vec{q}^2J^T_{11}-3 A_{c4d}q_0J^T_{21}+ (2A_{c4c}+A_{c4d})q_0\vec{q}^2J^T_{22} \nonumber \\
 && \quad +2 A_{c4c}\vec{q}^2J^T_{24}-6A_{c4c}q_0J^T_{31}+2A_{c4c}q_0\vec{q}^2J^T_{32}+2A_{c4c}\vec{q}^2J^T_{33} \nonumber\\
 && \quad -6A_{c4c}J^T_{34}  \bigg] \varepsilon(p_2)\cdot\varepsilon^*(p_4) \nonumber \\
 && \textnormal{with}  \;    \omega = v \cdot p_1 - M-\delta ; \\
&&\mathcal{M}^{(2)}_{(c5)} \nonumber\\
 &&= \frac{i}{4}\frac{g^4}{f^4} A_{c5}\bigg[ \Big (-\vec{q^2}J^B_{31}+5J^B_{41}-\vec{q}^2J^B_{42} \Big)\varepsilon(p_2)\cdot\varepsilon^*(p_4) + \Big(J^B_{21} \nonumber \\
 && \quad -\vec{q^2}J^B_{22}+7J^B_{31}-2\vec{q}^2J^B_{32}+7J^B_{42}-\vec{q}^2J^B_{43} \Big)   \nonumber \\
 && \quad \times q\cdot\varepsilon(p_2) q\cdot\varepsilon^*(p_4)  \bigg] \nonumber \\
 && \textnormal{with}  \;    \omega_1 = v \cdot p_1 - M-\delta, \; \omega_2 = v \cdot p_2 - M ; \\
&&\mathcal{M}^{(2)}_{(c6)} \nonumber\\
 &&= \frac{-i}{4}\frac{g^4}{f^4} A_{c6} \bigg[ \Big (-\vec{q^2}J^B_{21}+(\vec{q}^2)^2J^B_{22}-9\vec{q}^2J^B_{31}+2(\vec{q}^2)^2J^B_{32} \nonumber \\
 && \quad +10J^B_{41}-9\vec{q}^2J^B_{42}+(\vec{q}^2)^2J^B_{43} \Big)\varepsilon(p_2)\cdot\varepsilon^*(p_4) + \Big(-J^B_{21}\nonumber \\
 && \quad +\vec{q}^2J^B_{22}-7J^B_{31} +2\vec{q}^2J^B_{32}-7J^B_{42}+\vec{q}^2J^B_{43} \Big) \nonumber \\
 && \quad \times  q\cdot\varepsilon(p_2) q\cdot\varepsilon^*(p_4)  \bigg]    \nonumber \\
 && \textnormal{with}  \;    \omega_1 = v \cdot p_1 - M-\delta, \; \omega_2 = v \cdot p_2 - M-\delta ; \\
&&\mathcal{M}^{(2)}_{(c7)} \nonumber\\
 &&= \frac{-i}{4}\frac{g^4}{f^4} A_{c7} \bigg[ \vec{p}^2 J^B_{21} \varepsilon(p_2)\cdot\varepsilon^*(p_4) + J^B_{21} p\cdot\varepsilon(p_2) p\cdot\varepsilon^*(p_4) \bigg] \nonumber \\
 && \textnormal{with}  \;    \omega_1 = v \cdot p_1 - M -\delta, \; \omega_2 = v \cdot p_2 - M -\delta;  \\
&&\mathcal{M}^{(2)}_{(c8)} \nonumber\\
 &&= \frac{i}{4}\frac{g^4}{f^4} A_{c8} \bigg[ \Big( -\vec{q^2}J^R_{31}+5J^R_{41}-\vec{q}^2J^R_{42} \Big)\varepsilon(p_2)\cdot\varepsilon^*(p_4) + \Big(J^R_{21} \nonumber \\
 && \quad -\vec{q}^2J^R_{22}+7J^R_{31}-2\vec{q}^2J^R_{32}+7J^R_{42}-\vec{q}^2J^R_{43} \Big)  \nonumber \\
 && \quad \times q\cdot\varepsilon(p_2) q\cdot\varepsilon^*(p_4)  \bigg] \nonumber\\
 && \textnormal{with}  \;    \omega_1 = v \cdot p_1 - M-\delta, \; \omega_2 = v \cdot p_2 ; \\
&&\mathcal{M}^{(2)}_{(c9)} \nonumber\\
 &&= \frac{-i}{4}\frac{g^4}{f^4}  A_{c9}\bigg[ \Big (-\vec{q^2}J^R_{21}+(\vec{q}^2)^2J^R_{22}-9\vec{q}^2J^R_{31}+2(\vec{q}^2)^2J^R_{32} \nonumber \\
 && \quad +10J^R_{41} -9\vec{q}^2J^R_{42}+(\vec{q}^2)^2J^R_{43} \Big)\varepsilon(p_2)\cdot\varepsilon^*(p_4) + \Big(-J^R_{21} \nonumber \\
 && \quad +\vec{q}^2J^R_{22}-7J^R_{31} +2\vec{q}^2J^R_{32}-7J^R_{42}+\vec{q}^2J^R_{43} \Big) \nonumber \\
 && \quad \times q\cdot\varepsilon(p_2) q\cdot\varepsilon^*(p_4)  \bigg] \nonumber \\
 && \textnormal{with}  \;    \omega_1 = v \cdot p_1 - M-\delta, \; \omega_2 = v \cdot p_2 - M-\delta . \\
&&\mathcal{M}^{(2)}_{(c10)} \nonumber\\
 &&= \frac{i}{4}\frac{g^4}{f^4} A_{c10} \bigg[ \vec{p}^2 J^R_{21} \varepsilon(p_2)\cdot\varepsilon^*(p_4) +J^R_{21} p\cdot\varepsilon(p_2) p\cdot\varepsilon^*(p_4) \bigg] \nonumber \\
 && \textnormal{with}  \;    \omega_1 = v \cdot p_1 - M -\delta, \; \omega_2 = v \cdot p_2 - M -\delta.  \label{AllAmplitude2}
\end{eqnarray}

In above, $J^F_{ij}$ is the short notation for $J^F_{ij}(m_1,m_2,q)$, $J^S_{ij}$ and
 $J^T_{ij}$ are $J^S_{ij}(m_1,m_2,\omega,q)$ and $J^T_{ij}(m_1,m_2,\omega,q)$, respectively. $J^B_{ij}$ and
 $J^R_{ij}$ are $J^B_{ij}(m_1,m_2,\omega_1,\omega_2,q)$ and $J^R_{ij}(m_1,m_2,\omega_1,\omega_2,q)$, respectively.
These loop functions like $J^g$ are defined in the Appendix \ref{SecAppB}.

In Eqs.~(\ref{AllAmplitude1})-(\ref{AllAmplitude2}), the constants $A$ are different with different isospin. We list them in Tables \ref{TabContactA}, \ref{Tab1piA} and \ref{Tab2piA}.
The remaining constants are:
\begin{eqnarray}
A_{c1b}&=&1, \quad A_{c1c}=q_0^2, \quad A_{c2c}=-1, \quad  A_{c2d}=-1, \nonumber \\
A_{c3c}&=&-1, \quad A_{c3d}=-1, \quad A_{c4c}=1, \quad A_{c4d}=1,
\end{eqnarray}
for $I=1$. And
\begin{eqnarray}
A_{c1b}&=&-3, \quad A_{c1c}=-3q_0^2, \quad A_{c2c}=3, \quad  A_{c2d}=3, \nonumber \\
A_{c3c}&=&3, \quad A_{c3d}=3, \quad A_{c4c}=-3, \quad A_{c4d}=-3,
\end{eqnarray}
for $I=0$.
\begin{table}[!htbp]
	 \centering
	\caption{The coefficients for the contact amplitudes in the processes $DD^{*}\to DD^{*}$.
	}\label{TabContactA}
	\begin{tabular}{ccc}
		\hline
		   &   $I=1$   &    $I=0$  \\
		\hline
		$A_{a1}$   &   $ -32D_a-32E_a$         &   $-48D_a-48E_a$   \\
		$A_{a2}$   &   $8D_b-24D_a+8E_a+40E_b$  &   $24D_b-24D_a-24E_a+24E_b$     \\
		$A_{a3}$   &   $-32D_b-32E_b$          &   $48D_b+48E_b$    \\
		$A_{a4}$   &   $16D_b-80D_a-16E_a+80E_b$  &   $48D_b-96D_a-96E_a+48E_b$  \\
		$A_{a5}$   &   $-32D_b-32E_b$          &   $48D_b+48E_b$   \\
		$A_{a6}$   &   $8D_a-8D_b+8E_a-8E_b$   &   $24D_a+24D_b-72E_a-72E_b$     \\
		$A_{a7}$   &   $8D_a-8D_b+8E_a-8E_b$   &   $24D_a+24D_b-72E_a-72E_b$    \\
		$A_{a8}$   &   $0$  &   $-48D_b+144E_b$  \\
		$A_{a9}$   &   $0$    &   $-48D_b+144E_b$   \\
		$A_{a10}$   &  $16D_b-48E_b$           &   $0$     \\
		$A_{a11}$   &  $16D_b-48E_b$           &   $0$    \\
		$A_{a12a13}$   &   $-8D_a+8D_b-8E_a+8E_b$  &   $-8D_a-8D_b+24E_a+24E_b$  \\
		$A_{a14}$   &   $-8D_a+8D_b-8E_a+8E_b$  &   $-8D_a-8D_b+24E_a+24E_b$   \\
		\hline
	\end{tabular}
\end{table}

\begin{table}[!htbp]
	\centering
	\caption{The coefficients for the OPE amplitudes in the processes $DD^{*}\to D D^{*}$.
	}\label{Tab1piA}
	\begin{tabular}{ccccccccccc}
		\hline
		&   $A_{b1}$ &  $A_{b2}$ & $A_{b3}$ & $A_{b4}$ & $A_{b5}$ & $A_{b6}$ & $A_{b7}$ & $A_{f}$ & $A_{b10b11}$ & $A_{b12}$ \\
		\hline
		$I=1$ &   $-1$  &  $1$ & $-1$ & $1$ &   $1/3$ &   $1/3$ &   $-1$ &   $-1$ &   $-1$ &   $-1$  \\
		$I=0$ &   $-3$  &  $3$ & $-3$ & $3$ &   $1$ &   $1$ &   $-3$ &   $-3$ &   $-3$ &   $-3$  \\
		\hline
	\end{tabular}
\end{table}

\begin{table}[!htbp]
	\centering
	\caption{The coefficients for the TPE amplitudes in the processes $DD^{*}\to DD^{*}$.
	}\label{Tab2piA}
	\begin{tabular}{ccccccccccc}
		\hline
		&   $A_{c1a}$ &  $A_{c2}$ & $A_{c3}$ & $A_{c4}$ & $A_{c5}$ & $A_{c6}$ & $A_{c7}$ & $A_{c8}$ & $A_{c9}$ & $A_{c10}$ \\
		\hline
		$I=1$ &   $1$  &  $-2$ & $2$ & $-2$ &   $1$ &   $-1$ &   $-1$ &   $5$ &   $-5$ &   $-5$  \\
		$I=0$ &   $-3$  &  $-2$ & $2$ & $-2$ &   $9$ &   $-9$ &   $9$ &   $-3$ &   $3$ &   $-3$  \\
		\hline
	\end{tabular}
\end{table}

In Eqs.~(\ref{AllAmplitude1})-(\ref{AllAmplitude2}), $M$ is the $D$ meson mass, $\delta$ is the mass difference between $D^{*}$ and $D$, $m,m_1,m_2$ are all pion masses, $p=p_1-p_4$, $q=p_1-p_3$,
$\mu$ is the renormalization scale in the dimensional regularization, and
\begin{eqnarray}\label{DivergencePart}
L=\frac{1}{16\pi^2}\bigg(\frac{1}{d-4}+\frac{1}{2} \big(\gamma_E-1-\log 4\pi \big)\bigg).
\end{eqnarray}

\section{Renormalized and bare couplings}\label{secAppRBC}

We provide the relation between the renormalized coupling $g^{(2)}$ at experiment and the bare coupling $g$ in the Lagrangian
\begin{eqnarray} \label{gCorrection}
g^{(2)}&=& g \Bigg(1-\frac{g^2}{2f^2} J^g_{22}(0,-\delta)+\frac{g^2}{4f^2} J^g_{22}(-\delta,\delta) \nonumber \\
&& -\frac{9g^2}{8f^2}\partial J^b_{22}(-\delta) -\frac{3g^2}{8f^2}\partial J^b_{22}(\delta)-\frac{3g^2}{4f^2}\partial J^b_{22}(0)\Bigg).
\end{eqnarray}

The expression relating the renormalized $f_{(2)}$ and the bared $f$ is well known
\begin{eqnarray}
\frac{1}{f_{(2)}^2}=\frac{1}{f^2} \bigg(1+ \frac{m^2}{4\pi f^2} \textnormal{log}(\frac{m}{\mu}) \bigg).
\end{eqnarray}
We use $f_{(2)}=f_\pi=0.092$ GeV.

\section{Definitions of some loop functions} \label{SecAppB}
We define the loop functions following Ref.~\cite{Liu:2012vd}:
\begin{widetext}
\begin{eqnarray}
&& i\int\frac{d^D l \mu^{4-D} }{ {(2\pi)}^D } \frac{ \{1,~
	l^\alpha,~ l^\alpha l^\beta,~ l^\alpha l^\beta l^\gamma\} }
{[(+/-)v\cdot l+\omega+i\varepsilon](l^2-m^2+i\varepsilon)}
\nonumber\\&& \equiv \left\{J^{a/b}_0,~ v^\alpha J^{a/b}_{11}, ~
v^\alpha v^\beta J^{a/b}_{21}+g^{\alpha\beta}J^{a/b}_{22},~ (g\vee
v)J^{a/b}_{31}+v^\alpha v^\beta v^\gamma J^{a/b}_{32} \right\} (m,\omega),\label{LoopFunction1}
\\&&
i\int\frac{d^D l \mu^{4-D} }{ {(2\pi)}^D } \frac{ \{1,~
	l^\alpha,~ l^\alpha l^\beta,~ l^\alpha l^\beta l^\gamma\} }
{(v\cdot l+\omega_1+i\varepsilon)[(+/-)v\cdot
	l+\omega_2+i\varepsilon](l^2-m^2+i\varepsilon) }
\nonumber\\&&\equiv \left\{J^{g/h}_0,~ v^\alpha J^{g/h}_{11},
~ v^\alpha v^\beta
J^{g/h}_{21}+g^{\alpha\beta}J^{g/h}_{22},~ (g\vee
v)J^{g/h}_{31}+v^\alpha v^\beta v^\gamma J^{g/h}_{32} \right\}
(m,\omega_1,\omega_2),
\\&&
i\int\frac{d^D l \mu^{4-D} }{ {(2\pi)}^D } \frac{ \{1,~
	l^\alpha,~ l^\alpha l^\beta,~ l^\alpha l^\beta l^\gamma\} }
{(l^2-m_1^2+i\varepsilon)[(q+l)^2-m_2^2+i\varepsilon] }
\nonumber\\&&\equiv \left\{ J^F_0,~ q^\alpha J^F_{11},~
q^\alpha q^\beta J^F_{21}+g^{\alpha\beta}J^F_{22},~ (g\vee
q)J^F_{31}+q^\alpha q^\beta q^\gamma J^F_{32} \right\}(m_1,m_2,q),
\\&&
 i\int\frac{d^D l \mu^{4-D} }{ {(2\pi)}^D } \frac{ \{1,~
 	l^\alpha,~ l^\alpha l^\beta,~ l^\alpha l^\beta l^\gamma,~
 	l^\alpha l^\beta l^\gamma l^\delta\} } {[(+/-)v\cdot
 	l+\omega+i\varepsilon](l^2-m_1^2+i\varepsilon)[(q+l)^2-m_2^2+i\varepsilon]
 } \nonumber\\&&\equiv \left\{ J^{T/S}_0,~ q^\alpha
 J^{T/S}_{11}+v^\alpha J^{T/S}_{12},~ g^{\alpha \beta}
 J^{T/S}_{21}+q^\alpha q^\beta J^{T/S}_{22}+v^\alpha v^\beta
 J^{T/S}_{23}+(q\vee v)J^{T/S}_{24},\right.  (g\vee
 q)J^{T/S}_{31}+q^\alpha q^\beta q^\gamma J^{T/S}_{32} +(q^2\vee
 v)J^{T/S}_{33} \nonumber\\ && +(g\vee v)J^{T/S}_{34}+(q\vee v^2)J^{T/S}_{35}+v^\alpha v^\beta
 v^\gamma J^T_{36},~ (g\vee g)J^{T/S}_{41}+(g\vee
 q^2)J^{T/S}_{42}+q^\alpha q^\beta q^\gamma q^\delta J^{T/S}_{43} +(g\vee
 v^2)J^{T/S}_{44} + v^\alpha v^\beta v^\gamma v^\delta J^{T/S}_{45} \nonumber\\&&
 +(q^3\vee v) J^{T/S}_{46} +(q^2\vee v^2)J^{T/S}_{47} +(q\vee v^3) J^{T/S}_{48}
 \left. +(g\vee q\vee v) J^{T/S}_{49}\right\}(m_1,m_2,\omega,q),
\end{eqnarray}
\begin{eqnarray}
&&
i\int\frac{d^D l \mu^{4-D} }{ {(2\pi)}^D } \frac{ \{1,~
	l^\alpha,~ l^\alpha l^\beta,~ l^\alpha l^\beta l^\gamma,~
	l^\alpha l^\beta l^\gamma l^\delta\} } {(v\cdot
	l+\omega_1+i\varepsilon)[(+/-)v\cdot
	l+\omega_2+i\varepsilon](l^2-m_1^2+i\varepsilon)[(q+l)^2-m_2^2+i\varepsilon]
} \nonumber\\&& \equiv \left\{ J^{R/B}_0,~ q^\alpha
J^{R/B}_{11}+v^\alpha J^{R/B}_{12},~ g^{\alpha \beta}
J^{R/B}_{21}+q^\alpha q^\beta J^{R/B}_{22}+v^\alpha v^\beta
J^{R/B}_{23}+(q\vee v)J^{R/B}_{24},\right.  (g\vee
q)J^{R/B}_{31}+q^\alpha q^\beta q^\gamma J^{R/B}_{32} +(q^2\vee
v)J^{R/B}_{33} \nonumber\\ &&  +(g\vee v)J^{R/B}_{34} +(q\vee
v^2)J^{R/B}_{35}+v^\alpha v^\beta v^\gamma J^{R/B}_{36},
(g\vee g)J^{R/B}_{41}+(g\vee
q^2)J^{R/B}_{42}+q^\alpha q^\beta q^\gamma q^\delta J^{R/B}_{43}
+(g\vee v^2)J^{R/B}_{44} + v^\alpha v^\beta v^\gamma v^\delta
J^{R/B}_{45} \nonumber\\&& +(q^3\vee v) J^{R/B}_{46}+(q^2\vee v^2)J^{R/B}_{47}
 \left.+(q\vee v^3) J^{R/B}_{48}+(g\vee q\vee v)
J^{R/B}_{49}
\right\}(m_1,m_2,\omega_1,\omega_2,q),  \label{LoopFunction2}
\end{eqnarray}
with
\begin{eqnarray}
&&q \vee v \equiv q^\alpha v^\beta+q^\beta v^\alpha, \quad g \vee
q \equiv
g^{\alpha\beta}q^\gamma+g^{\alpha\gamma}q^\beta+g^{\gamma\beta}q^\alpha,
\quad g \vee v \equiv
g^{\alpha\beta}v^\gamma+g^{\alpha\gamma}v^\beta+g^{\gamma\beta}v^\alpha,
\quad \nonumber\\&& q^2 \vee v \equiv q^{\beta } q^{\gamma }
v^{\alpha }+q^{\alpha }
q^{\gamma } v^{\beta }+q^{\alpha } q^{\beta } v^{\gamma }, \quad
q \vee v^2 \equiv q^{\gamma } v^{\alpha }
v^{\beta }+q^{\beta } v^{\alpha } v^{\gamma }+q^{\alpha } v^{\beta } v^{\gamma },\quad
\nonumber\\&& g \vee g \equiv g^{\alpha \beta } g^{\gamma \delta
}+g^{\alpha \delta } g^{\beta \gamma }+g^{\alpha \gamma } g^{\beta
\delta }, \quad g \vee q^2 \equiv q^{\alpha } q^{\beta } g^{\gamma
\delta }+q^{\alpha } q^{\delta } g^{\beta \gamma } +q^{\alpha}
q^{\gamma } g^{\beta \delta }+q^{\gamma } q^{\delta } g^{\alpha
	\beta } +q^{\beta } q^{\delta } g^{\alpha \gamma } +q^{\beta }
q^{\gamma } g^{\alpha \delta }, \quad \nonumber\\&& g \vee v^2
\equiv v^{\alpha } v^{\beta } g^{\gamma \delta } +v^{\alpha }
v^{\delta } g^{\beta \gamma }+v^{\alpha } v^{\gamma } g^{\beta
	\delta }
+v^{\gamma } v^{\delta } g^{\alpha \beta }+v^{\beta } v^{\delta   } g^{\alpha \gamma }
+v^{\beta } v^{\gamma } g^{\alpha \delta }, \quad
\nonumber\\&& q^3\vee v \equiv q^{\beta } q^{\gamma } q^{\delta }
v^{\alpha }+q^{\alpha } q^{\gamma } q^{\delta} v^{\beta }
+q^{\alpha } q^{\beta } q^{\delta } v^{\gamma }+q^{\alpha }
q^{\beta } q^{\gamma } v^{\delta } ,\quad q\vee v^3 \equiv
q^{\delta } v^{\alpha } v^{\beta } v^{\gamma }+q^{\gamma }
v^{\alpha } v^{\beta } v^{\delta } +q^{\beta } v^{\alpha }
v^{\gamma } v^{\delta }+q^{\alpha } v^{\beta } v^{\gamma }
v^{\delta }, \nonumber\\&& q^2 \vee v^2 \equiv q^{\gamma }
q^{\delta } v^{\alpha } v^{\beta }+q^{\beta } q^{\delta }
v^{\alpha } v^{\gamma } +q^{\alpha } q^{\delta } v^{\beta }
v^{\gamma }+q^{\beta } q^{\gamma } v^{\alpha } v^{\delta }
+q^{\alpha } q^{\gamma } v^{\beta }  v^{\delta }+q^{\alpha }
q^{\beta } v^{\gamma } v^{\delta }, \nonumber\\&& g\vee q \vee v
\equiv q^{\beta } v^{\alpha } g^{\gamma \delta }+q^{\alpha }
v^{\beta } g^{\gamma \delta } +q^{\delta } v^{\alpha } g^{\beta
	\gamma }+q^{\gamma } v^{\alpha } g^{\beta \delta }+q^{\alpha }
v^{\delta } g^{\beta \gamma } +q^{\alpha } v^{\gamma } g^{\beta
	\delta}+q^{\delta } v^{\gamma } g^{\alpha \beta }+q^{\delta }
v^{\beta } g^{\alpha \gamma } +q^{\gamma } v^{\delta } g^{\alpha
	\beta  } \nonumber\\&& \qquad\qquad\quad+q^{\gamma } v^{\beta }
g^{\alpha \delta }+q^{\beta } v^{\delta } g^{\alpha \gamma }
+q^{\beta } v^{\gamma } g^{\alpha \delta }.
\end{eqnarray}
\end{widetext}

$J^b$ is related to $J^a$:
 \begin{eqnarray}
 && J^b_0=J^a_0, \quad J^b_{11}=-J^a_{11}, \quad J^b_{21}=J^a_{21}, \quad J^b_{22}=J^a_{22}, \nonumber \\
 && J^b_{31}=-J^a_{31}, \quad J^b_{32}=-J^a_{32}.
 \end{eqnarray}
$J^g$ and $J^h$ can be deduced to
 \begin{eqnarray}
 && J^g(\omega_1,\omega_2) =  \frac{1}{\omega_2-\omega_1} \left[ J^a(\omega_1)-J^a(\omega_2)\right],  \\
 && J^h(\omega_1,\omega_2) =  \frac{1}{\omega_2+\omega_1} \left[ J^a(\omega_1)+J^b(\omega_2)\right].
 \end{eqnarray}
$J^S$ is related to $J^T$:
\begin{eqnarray}
&& J^S_0(v \cdot q) = J^T_0(-v \cdot q) , \quad J^S_{11}(v\cdot q)=J^T_{11}(-v\cdot q), \nonumber \\
&& J^S_{12}(v \cdot q)=-J^T_{12}(-v\cdot q), \quad J^S_{21}=J^T_{21}(-v\cdot q), \nonumber \\
&& J^S_{22}(v \cdot q)=J^T_{22}(-v\cdot q), \quad J^S_{23}(v \cdot q)=J^T_{23}(-v \cdot q ). \nonumber \\
&& J^S_{24}(v \cdot q)=-J^T_{24}(-v\cdot q), \quad J^S_{31}(v \cdot q)=J^T_{31}(-v \cdot q ). \nonumber \\
&& J^S_{32}(v \cdot q)=J^T_{32}(-v\cdot q), \quad J^S_{33}(v \cdot q)=-J^T_{33}(-v \cdot q ). \nonumber \\
&& J^S_{34}(v \cdot q)=-J^T_{34}(-v\cdot q), \quad J^S_{35}(v \cdot q)=J^T_{35}(-v \cdot q ). \nonumber \\
&& J^S_{36}(v \cdot q)=-J^T_{34}(-v\cdot q), \quad J^S_{41}(v \cdot q)=J^T_{41}(-v \cdot q ). \nonumber \\
&& J^S_{42}(v \cdot q)=J^T_{42}(-v\cdot q), \quad J^S_{43}(v \cdot q)=J^T_{43}(-v \cdot q ). \nonumber \\
&& J^S_{44}(v \cdot q)=J^T_{44}(-v\cdot q), \quad J^S_{45}(v \cdot q)=J^T_{45}(-v \cdot q ). \nonumber \\
&& J^S_{46}(v \cdot q)=-J^T_{46}(-v\cdot q), \quad J^S_{47}(v \cdot q)=J^T_{47}(-v \cdot q ). \nonumber \\
&& J^S_{48}(v \cdot q)=-J^T_{48}(-v\cdot q), \quad J^S_{49}(v \cdot q)=-J^T_{49}(-v \cdot q ). \nonumber\\
\end{eqnarray}
$J^R$ and $J^B$ can be deduced to
\begin{eqnarray}
&& J^R(\omega_1,\omega_2) =  \frac{1}{\omega_2-\omega_1} \left[ J^T(\omega_1)-J^T(\omega_2)\right],  \\
&& J^B(\omega_1,\omega_2) =  \frac{1}{\omega_2+\omega_1} \left[ J^T(\omega_1)+J^S(\omega_2)\right].
\end{eqnarray}

All the integrals in Eqs.~(\ref{LoopFunction1})-(\ref{LoopFunction2}) can be deduced to one or two dimensional
Feynmann parameter integrals without difficulty. For example,
\begin{eqnarray}
&& J^T_{36}  \nonumber \\
&& =2L\int_{0}^{1}dx_1(4b^2-c)+\frac{3}{16\pi^2}\int_{0}^{1}dx_1b^2  +\frac{1}{16\pi^2} \int_0^1dx_1 \nonumber \\
&& \quad  \times(4b^2-c)\left[-\log \mu^2  +\log (-b^2+c) \right] -\frac{3}{16\pi}\int^1_0dx_1b \nonumber\\
&& \quad \times (-b^2+c)^{\frac{1}{2}}+\frac{1}{16\pi} \int^1_0dx_1b^3(-b^2+c)^{-\frac{1}{2}} \nonumber \\
&& \quad -\frac{1}{8\pi^2}\int^1_0 dx_1 D,
\end{eqnarray}
\begin{eqnarray}
&& J^T_{45}  \nonumber \\
&& =8L\int_{0}^{1}dx_1b(2b^2-c)+\frac{1}{4\pi^2}\int_{0}^{1}dx_1b^3  +\frac{1}{4\pi^2} \int_0^1dx_1 \nonumber \\
&& \quad  \times b(2b^2-c)\left[-\log \mu^2  +\log (-b^2+c) \right] +\frac{1}{16\pi}\int^1_0dx_1 \nonumber\\
&& \quad \times (-b^2+c)^{\frac{3}{2}}-\frac{3}{8\pi} \int^1_0dx_1b^2(-b^2+c)^{\frac{1}{2}}+\frac{1}{16\pi}\int^1_0dx_1 \nonumber \\
&& \quad \times b^4(-b^2+c)^{-\frac{1}{2}}+\frac{1}{8\pi^2}\int^1_0 dx_1 E,
\end{eqnarray}
where
\begin{eqnarray}
b&=& (1-x_1)v\cdot q-\omega, \nonumber \\
c &=& (1-x_1)^2q^2-(1-x_1)q^2+x_1(m^2_1-m^2_2)+m^2_2-i\epsilon, \nonumber \\
D&=&\left\{ \sqrt{c-b^2} \left[ (4b^2-c)\log \left(1-\frac{b^2}{c} \right)+5b^2 \right] + (8b^3-6bc) \right. \nonumber \\
&& \left. \times \tan^{-1}\left(\frac{b}{\sqrt{c-b^2}}\right) \right\}\left(2\sqrt{c-b^2}\right)^{-1}, \nonumber \\
E&=&\Bigg\{ b\sqrt{c-b^2} \bigg[ 6(2b^2-c) \bigg( \log(c)-\log \left[c-b^2 \right]\bigg)-16b^2   \nonumber \\
&&  +3c \bigg] -3(8b^4-8b^2c+c^2)\tan^{-1}\left(\frac{b}{\sqrt{c-b^2}}\right) \Bigg\}  \nonumber \\
&& \times \left(3\sqrt{c-b^2}\right)^{-1}, \nonumber \\
\end{eqnarray}
and $L$ is defined in Eq.~(\ref{DivergencePart}).

One should notice that in Eqs.~(\ref{LoopFunction1})-(\ref{LoopFunction2}), if the form of the integral
Eq.~(\ref{DiagramWith2PR}) is encountered, the 2PR part must be subtracted using Eq.~(\ref{SubtractionOf2PR}).

However, the evaluations of above loop integrals are not complete since the kinetic energy terms in the propagators are not included. Here, we further illustrate the calculations considering the kinetic energy terms $\frac{\vec{q^2}}{2M}$. We choose $J^b_0$ as an example,
\begin{eqnarray} \label{Jb0q2}
i\int\frac{d^Dl \mu^{4-D}}{(2\pi)^D}\frac{1}{\Big[-v\cdot l- \frac{(\vec{p}-\vec{l})^2}{2M}+\omega+i\varepsilon\Big]\big[l^2-m^2+i\varepsilon\big]}.
\end{eqnarray}

We first apply the Feynman parametrization to Eq.~(\ref{Jb0q2}):
\begin{widetext}
\begin{eqnarray}
\frac{1}{\Big[-v\cdot l- \frac{(\vec{p}-\vec{l})^2}{2M}+\omega+i\varepsilon\Big]\big[l^2-m^2+i\varepsilon\big]}
&=&2\int^\infty_0 dy \frac{1}{\Big[l^2-m^2+2y\Big(-v\cdot l-\frac{(\vec{p}-\vec{l})^2}{2M}+\omega\Big)+i\varepsilon\Big]^2}\nonumber \\
&=&2\int^\infty_0 dy \frac{1}{\Big[l^2-2yv\cdot l +y^2v^2-y^2v^2-\frac{y}{M}(\vec{p}-\vec{l})^2+2y\omega-m^2+i\varepsilon\Big]^2}\nonumber \\
&=&2\int^\infty_0 dy \frac{1}{\Big[(l-yv)^2-y^2-\frac{y}{M}(\vec{p}-\vec{l})^2+2y\omega-m^2+i\varepsilon\Big]^2}.
\end{eqnarray}
\end{widetext}  	
with the substitution $l\to l+yv$ we obtain
\begin{eqnarray}
2\int^\infty_0 dy \frac{1}{\Big[l^2-y^2-\frac{y}{M}(\vec{p}-\vec{l})^2+2y\omega-m^2+i\varepsilon\Big]^2}.
\end{eqnarray}

Next, we analyze the pole structure of the expression and perform $l_0$ integral. We first rewrite
the polynomial of $l_0$ in the denominator:
\begin{eqnarray}
&&l^2-y^2-\frac{y}{M}(\vec{p}-\vec{l})^2+2y\omega-m^2+i\varepsilon\nonumber \\
&&=l^2_0-\vec{l}^2-y^2-\frac{y}{M}(\vec{p}-\vec{l})^2+2y\omega-m^2+i\varepsilon\nonumber \\
&&=l^2_0-\Big[\vec{l}^2+\frac{y}{M}(\vec{p}-\vec{l})^2+y^2-2y\omega+m^2\Big]+i\varepsilon\nonumber \\
&&=\big[l_0+E_l\big]\big[l_0-E_l\big],
\end{eqnarray}
where $E_l=\sqrt{\vec{l}^2+\frac{y}{M}(\vec{p}-\vec{l})^2+y^2-2y\omega+m^2}-i\varepsilon$. Therefore there exist
 two poles located at $-E_l$ and $E_l$.
 
With the expressions above, Eq.~(\ref{Jb0q2}) becomes
\begin{eqnarray}\label{Jb0q21}
&&i\int\frac{d^Dl \mu^{4-D}}{(2\pi)^D}\frac{1}{\Big[-v\cdot l- \frac{(\vec{p}-\vec{l})^2}{2M}+\omega+i\varepsilon\Big]\big[l^2-m^2+i\varepsilon\big]}\nonumber \\
&&=2i\int^\infty_0 dy\int\frac{d^{D}l \mu^{4-D}}{(2\pi)^D}\frac{1}{\big[l_0+E_l\big]^2\big[l_0-E_l\big]^2}\nonumber \\
&&=2i\int^\infty_0 dy\int\frac{d^{D-1}l  \mu^{4-D}}{(2\pi)^D}\int dl_0\frac{1}{\big[l_0+E_l\big]^2\big[l_0-E_l\big]^2}.\nonumber\\
\end{eqnarray}

By closing the contour in the upper complex $l_0$ plane, we obtain the $l_0$ integral 
\begin{eqnarray}
\int dl_0\frac{1}{\big[l_0+E_l\big]^2\big[l_0-E_l\big]^2}=2\pi i\textnormal{Res}\big(f(-E_l)\big),
\end{eqnarray}
where $\textnormal{Res}\big(f(-E_l)\big)$ is the residue at $-E_l$, it can be evaluated using
\begin{eqnarray}
\textnormal{Res}(f(z_0))=\lim_{z\to z_0}\frac{1}{(m-1)!}\bigg\{\frac{d^{m-1}}{dz^{m-1}}\big[(z-z_0)^mf(z)\big]\bigg\},
\end{eqnarray}
i.e.,
\begin{eqnarray}
&&\textnormal{Res}\big(f(-E_l)\big)\nonumber\\
&&=\lim_{l_0\to -E_l}\bigg\{\frac{d}{dl_0}\bigg[\Big(l_0-(-E_l)\Big)^2\frac{1}{\big[l_0+E_l\big]^2\big[l_0-E_l\big]^2}\bigg]\bigg\}\nonumber\\
&&=\lim_{l_0\to -E_l}\frac{2}{(E_l-l_0)^3}\nonumber\\
&&=\frac{2}{(2E_l)^3}\nonumber\\
&&=\frac{1}{4}\frac{1}{\big[\vec{l}^2+\frac{y}{M}(\vec{p}-\vec{l})^2+y^2-2y\omega+m^2-i\varepsilon\big]^{3/2}},
\end{eqnarray}
where the expression $\vec{l}^2+\frac{y}{M}(\vec{p}-\vec{l})^2+y^2-2y\omega+m^2-i\varepsilon$ should be further simplified:
\begin{eqnarray}
&&\bigg(1+\frac{y}{M}\bigg)\bigg[\vec{l}-\frac{\frac{y}{M}}{1+\frac{y}{M}}\vec{p}\bigg]^2+\frac{\frac{y}{M}}{1+\frac{y}{M}}\vec{p}^2
+(y-\omega)^2+m^2-\omega^2\nonumber\\
&&=\bigg(1+\frac{y}{M}\bigg)\vec{l}^2+(y-\omega)^2+m^2-\omega^2\nonumber\\
&&=\bigg(1+\frac{y}{M}\bigg)\bigg[\vec{l}^2+\frac{(y-\omega)^2+m^2-\omega^2}{1+\frac{y}{M}}\bigg]\nonumber\\
&&=\bigg(1+\frac{y}{M}\bigg)\big[\vec{l}^2+\Delta\big]
\end{eqnarray}
with
\begin{eqnarray}
\Delta=\frac{(y-\omega)^2+m^2-\omega^2}{1+\frac{y}{M}}.
\end{eqnarray}

Then, Eq.~(\ref{Jb0q21}) reduces to
\begin{eqnarray}\label{Jb0q22}
&&2i\int^\infty_0 dy\int\frac{d^{D-1}l  \mu^{4-D}}{(2\pi)^D}(2\pi i) \frac{1}{4} \frac{1}{\bigg[\bigg(1+\frac{y}{M}\bigg)\big[\vec{l}^2+\Delta\big]\bigg]^{3/2}}\nonumber\\
&&=-\frac12\int^\infty_0 dy\int\frac{d^{D-1}l  \mu^{4-D}}{(2\pi)^{D-1}} \frac{1}{\bigg[\bigg(1+\frac{y}{M}\bigg)\big[\vec{l}^2+\Delta\big]\bigg]^{3/2}}\nonumber\\
&&=-\frac12\int^\infty_0 dy\frac{\mu^{4-D}}{(2\pi)^{\frac{D-1}{2}}}\frac{\Gamma\Big[2-\frac{D}{2}\Big]}{\Gamma\big[\frac32\big]} \frac{1}{\bigg(1+\frac{y}{M}\bigg)^{3/2}\Delta^{2-\frac{D}{2}}}.
\end{eqnarray}    
Using
\begin{eqnarray}
y\to y+\omega, \quad \Delta\to \frac{ y^2+m^2-\omega^2}{1+\frac{y+\omega}{M}},
\end{eqnarray}
 Eq.~(\ref{Jb0q22}) can be further simplified
\begin{eqnarray}
&&-\frac12\int^\infty_{-\omega} dy\frac{\mu^{4-D}}{(2\pi)^{\frac{D-1}{2}}}\frac{\Gamma\Big[2-\frac{D}{2}\Big]}{\Gamma\big[\frac32\big]} \frac{1}{\bigg(1+\frac{y+\omega}{M}\bigg)^{3/2}\Delta^{2-\frac{D}{2}}}\nonumber\\
&&=-\frac12\int^\infty_{-\omega} dy\frac{\mu^{\epsilon}}{(2\pi)^{\frac{3-\epsilon}{2}}}\frac{\Gamma\Big[\frac{\epsilon}{2}\Big]}{\Gamma\big[\frac32\big]} \frac{1}{\bigg(1+\frac{y+\omega}{M}\bigg)^{3/2}\Delta^{\frac{\epsilon}{2}}}\nonumber\\
&&=\frac{-1}{2}\int^\infty_{0} dy\frac{\mu^{\epsilon}}{(2\pi)^{\frac{3-\epsilon}{2}}}\frac{\Gamma\Big[\frac{\epsilon}{2}\Big]}{\Gamma\big[\frac32\big]} \frac{1}{\bigg(1+\frac{y+\omega}{M}\bigg)^{3/2}\Delta^{\frac{\epsilon}{2}}}\nonumber\\
&&+\frac{-1}{2}\int^0_{-\omega} dy\frac{\mu^{\epsilon}}{(2\pi)^{\frac{3-\epsilon}{2}}}\frac{\Gamma\Big[\frac{\epsilon}{2}\Big]}{\Gamma\big[\frac32\big]} \frac{1}{\bigg(1+\frac{y+\omega}{M}\bigg)^{3/2}\Delta^{\frac{\epsilon}{2}}},
\end{eqnarray}
where $\epsilon=4-D$.

We first discuss the $\int^\infty_0$ part,
\begin{align}
&\frac{-1}{2}\int^\infty_{0} dy\frac{\mu^{\epsilon}}{(2\pi)^{\frac{3-\epsilon}{2}}}\frac{\Gamma\Big[\frac{\epsilon}{2}\Big]}{\Gamma\big[\frac32\big]} \frac{1}{\bigg(1+\frac{y+\omega}{M}\bigg)^{3/2}\Delta^{\frac{\epsilon}{2}}}\nonumber\\
&=\frac{-1}{2}\frac{(4\pi)^{\frac12}}{\Gamma\big[\frac32\big]}\frac{\mu^\epsilon\Gamma\Big[\frac{\epsilon}{2}\Big]}{(4\pi)^{2-\frac{\epsilon}{2}}}\int^\infty_{0} dy\frac{\bigg(1+\frac{y+\omega}{M}\bigg)^{\frac{-3}{2}}}{\big(\frac{y^2-\omega^2+m^2}{1+\frac{y+\omega}{M}}\big)^{\frac{\epsilon}{2}}}\nonumber\\
&=\frac{-1}{2}\frac{(4\pi)^{\frac12}}{\Gamma\big[\frac32\big]}\frac{\mu^\epsilon\Gamma\Big[\frac{\epsilon}{2}\Big]}{(4\pi)^{2-\frac{\epsilon}{2}}}\int^\infty_{0} dy\frac{\bigg(1+\frac{y+\omega}{M}\bigg)^{\frac{\epsilon-3}{2}}}{\big(y^2-\omega^2+m^2\big)^{\frac{\epsilon}{2}}}.
\end{align}
Notice that, if we assume $M\to\infty$, the expression above becomes
\begin{align}
&\frac{-1}{2}\frac{(4\pi)^{\frac12}}{\Gamma\big[\frac32\big]}\frac{\mu^\epsilon\Gamma\Big[\frac{\epsilon}{2}\Big]}{(4\pi)^{2-\frac{\epsilon}{2}}}\int^\infty_{0} dy\frac{1}{\big(y^2-\omega^2+m^2\big)^{\frac{\epsilon}{2}}} \nonumber\\
&=-2\frac{\mu^\epsilon\Gamma\Big[\frac{\epsilon}{2}\Big]}{(4\pi)^{2-\frac{\epsilon}{2}}}\frac{\Gamma\big[\frac12\big]\Gamma\big[\frac{-1}{2}\big]}{2\Gamma\Big[\frac{\epsilon}{2}\Big]}\big(-\omega^2+m^2\big)^{\frac12-\frac{\epsilon}{2}} \nonumber\\
&=\frac{1}{8\pi}\big(-\omega^2+m^2\big)^{\frac12}.
\end{align}
The result above reproduces part of $J^b_0$ where $\vec{q}^2/M$ in the propagator is not included.

We now discuss the $\int^0_{-\omega}$ part:
\begin{eqnarray}
&&\frac{-1}{2}\int^0_{-\omega} dy\frac{\mu^{\epsilon}}{(2\pi)^{\frac{3-\epsilon}{2}}}\frac{\Gamma\Big[\frac{\epsilon}{2}\Big]}{\Gamma\big[\frac32\big]} \frac{1}{\bigg(1+\frac{y+\omega}{M}\bigg)^{3/2}\Delta^{\frac{\epsilon}{2}}}\nonumber\\
&&=\frac{-1}{2}\frac{(4\pi)^{\frac12}}{\Gamma\big[\frac32\big]}\frac{\mu^\epsilon\Gamma\Big[\frac{\epsilon}{2}\Big]}{(4\pi)^{2-\frac{\epsilon}{2}}}\int^0_{-\omega} dy\frac{\bigg(1+\frac{y+\omega}{M}\bigg)^{\frac{\epsilon-3}{2}}}{\big(y^2-\omega^2+m^2\big)^{\frac{\epsilon}{2}}}\nonumber\\
&&=-2\bigg(-2L+\frac{1}{8\pi^2}\textnormal{log}\,\mu-\frac{1}{16\pi^2}\bigg)\int^0_{-\omega} dy\bigg(1+\frac{y+\omega}{M}\bigg)^{\frac{-3}{2}}\nonumber\\
&&\quad\times \frac{\bigg(1+\frac{y+\omega}{M}\bigg)^{\frac{\epsilon}{2}}}{\big(y^2-\omega^2+m^2\big)^{\frac{\epsilon}{2}}}\nonumber\\
&&=-2\bigg(-2L+\frac{1}{8\pi^2}\textnormal{log}\,\mu-\frac{1}{16\pi^2}\bigg)\int^0_{-\omega} dy\bigg(1+\frac{y+\omega}{M}\bigg)^{\frac{-3}{2}}\nonumber\\
&&\quad\times \Bigg[1+\frac{\epsilon}{2}\textnormal{log}\frac{1+\frac{y+\omega}{M}}{y^2-\omega^2+m^2}\Bigg]\nonumber\\
&&=\bigg(4L-\frac{1}{4\pi^2}\textnormal{log}\,\mu+\frac{1}{8\pi^2}\bigg)\int^0_{-\omega} dy\bigg(1+\frac{y+\omega}{M}\bigg)^{\frac{-3}{2}}\nonumber\\
&&\quad-\frac{1}{8\pi^2} \int^0_{-\omega} dy\bigg(1+\frac{y+\omega}{M}\bigg)^{\frac{-3}{2}}\textnormal{log}\frac{1+\frac{y+\omega}{M}}{y^2-\omega^2+m^2},
\end{eqnarray}
where the term containing $L$ (defined in Eq.~(\ref{DivergencePart})) is a divergent part. The expression above will be further evaluated numerically. 
If we assume $M\to\infty$ again, the result can reproduce another part of $J^b_0$ where $\vec{q}^2/M$ in the propagator is not included 
at the beginning.

The evaluations of other loop integrals in Eqs.~(\ref{LoopFunction1})-(\ref{LoopFunction2}) are similar.

\end{document}